\documentclass[journal,onecolumn,12pt,draftclsnofoot]{IEEEtran}
\usepackage{graphicx}
\usepackage{amsfonts}
\usepackage{amsmath}
\usepackage{amssymb}
\usepackage{bm}
\usepackage{color}
\usepackage{graphicx,subfigure}
\usepackage{amssymb}
\usepackage{times}
\usepackage{latexsym}
\usepackage{bm}
\usepackage{cases}
\usepackage{array}
\usepackage{subfigure}
\usepackage{url}
\usepackage{algorithm}
\usepackage{algorithmic}
\usepackage{multirow}
\usepackage{epstopdf}
\usepackage{epsfig}
\usepackage{fancybox}
\usepackage{textcomp}

\usepackage{cite}
\begin{document}



\title{Multi-hop RIS-Empowered Terahertz Communications: A DRL-based \\
Hybrid Beamforming Design}

\author{Chongwen~Huang,~\IEEEmembership{Member,~IEEE,}  Zhaohui~Yang,~\IEEEmembership{Member,~IEEE,} George~C.~Alexandropoulos,~\IEEEmembership{Senior Member,~IEEE}, Kai Xiong, Li Wei Chau~Yuen,~\IEEEmembership{Fellow,~IEEE}, Zhaoyang Zhang,~\IEEEmembership{Member,~IEEE}, and M\'{e}rouane~Debbah,~\IEEEmembership{Fellow,~IEEE}

\thanks{C. Huang and Z. Zhang are with the College of Information Science and Electronic Engineering, Zhejiang Provincial Key Lab of information processing, communication and networking, Zhejiang University, Hangzhou, 310007, P.R. China. They are also with the ZJU-UIUC Institute, Zhejiang University, Haining,
314400, China. (e-mails: chongwenhuang@zju.edu.cn, ning\_ming@zju.edu.cn). The work of Prof. Zhang was supported in part by the National Key R\&D Program of China under grant 2020YFB1807101 and the National Natural Science Foundation of China under Grant 61725104.}
\thanks{Z. Yang is with the Centre For Telecommunications Research, Department Of Engineering, King's college london, WC2R 2LS, UK. (e-mail: yang.zhaohui@kcl.ac.uk).}
\thanks{G.~C.~Alexandropoulos is with the Department of Informatics and Telecommunications, National and Kapodistrian University of Athens, Panepistimiopolis Ilissia, 15784 Athens, Greece. (e-mail: alexandg@di.uoa.gr).}
\thanks{K. Xiong is with School of Information and Communication Engineering, University of Electronic Science and Technology of China, Chengdu, 611731, China.
(e-mail: xiongkaipai@163.com). }
\thanks{Li Wei and C. Yuen are with the Singapore University of Technology and Design, 487372 Singapore. (emails: wei\_li@mymail.sutd.edu.sg, yuenchau@sutd.edu.sg).}

\thanks{M.~Debbah is with CentraleSup\'elec, University Paris-Saclay, 91192 Gif-sur-Yvette, France. M. Debbah is also with the Lagrange Mathematical and Computing Research Center,
Paris, 75007 France (email: merouane.debbah@huawei.com). }

}

\maketitle

\begin{abstract}

Wireless communication in the TeraHertz band (0.1--10 THz) is envisioned as one of the key enabling technologies for the future sixth generation (6G) wireless communication systems scaled up beyond massive multiple input multiple output (Massive-MIMO) technology. However, very high propagation attenuations and molecular absorptions of THz frequencies often limit the signal transmission distance and coverage range. Benefited from the recent breakthrough on the reconfigurable intelligent surfaces (RIS) for realizing smart radio propagation environment, we propose a novel hybrid beamforming scheme for the multi-hop RIS-assisted communication networks to improve the coverage range at THz-band frequencies. Particularly, multiple passive and controllable RISs are deployed to assist the transmissions between the base station (BS) and multiple single-antenna users. We investigate the joint design of digital beamforming matrix at the BS and analog beamforming matrices at the RISs, by leveraging the recent advances in deep reinforcement learning (DRL) to combat the propagation loss.  To improve the convergence of the proposed DRL-based algorithm, two algorithms are then designed to initialize the digital beamforming and the analog beamforming matrices utilizing the alternating optimization technique. Simulation results show that our proposed scheme is able to improve 50\% more coverage range of THz communications compared with the benchmarks. Furthermore, it is also shown that our proposed DRL-based method is a state-of-the-art method to solve the NP-hard beamforming problem, especially when the signals at RIS-assisted THz communication networks experience multiple hops.
\end{abstract}


\vspace{-0.25cm}
\section{Introduction}

Future sixth generation (6G) wireless communication systems are expected to rapidly evolve towards an  ultra-high speed and low latency with the software-based functionality paradigm \cite{Akyildiz_Thz2018wcm,rappaport2019,testbed2,sarieddeen2020overview,han2018mag,Sarieddeen2020wcm,han2019terahertz,liaskos2018,chongwenhmimos,Ref3a00,Ref3a01,Ref4a}.  Although current millimeter-wave (mmWave) communication systems (30-300 GHz) have been  integrated into 5G mobile systems, and several mmWave sub-bands were released for licensed communications, e.g., 57-64 GHz, 81-86 GHz, etc., the total consecutive available bandwidth is still less than 10 GHz, which is difficult to offer Tbps data rates\cite{Akyildiz_Thz2018wcm,rappaport2019,sarieddeen2020overview,Sarieddeen2020wcm,han2018mag,liaskos2018,han2019terahertz,Nie2020beamforming,han2016dis}. To meet the increasing demand for higher data rates and new spectral bands, the Terahertz (0.1--10 THz) band communication is considered as one of the promising technology to enable ultra-high speed and low-latency communications.

Although  major progresses in the recent ten years are empowering practical THz communication networks,  there are still  many challenges in THz communications that require innovative solutions. One of the major challenges is the very high propagation attenuations, which drastically reduces the propagation distance\cite{Akyildiz_Thz2018wcm,rappaport2019,sarieddeen2020overview,han2018mag,Sarieddeen2020wcm,han2019terahertz,liaskos2018}. To be specific, practical field experiments show that free space path attenuation of a 10-meter transmission distance at THz frequencies usually is larger than 100 dB\cite{Akyildiz_Thz2018wcm,rappaport2019,han2016dis,Moldovan2016,Nie2019resource}. On the one hand, it is ascribed to the spreading loss due to the quadratically increasing frequency. On the other hand, molecular absorption caused by water vapor and oxygen molecules also make considerable contributions to path loss. What's more, transmission power of THz transceivers limited by fabricated technologies is still in milli-Watt scale, which further reduces the propagation distance\cite{Akyildiz_Thz2018wcm,rappaport2019,han2016dis,Moldovan2017,ma2020intelligent}.

Fortunately, the recently proposed reconfigurable intelligent surface (RIS) is considered as a promising technology to combat the propagation distance problem, since RIS can be programmed to change an impinging electromagnetic (EM) field in a desired way to focus, steer, and enhance the signal power towards the object user \cite{liaskos2018,chongwenhmimos,chongwentwc2019,georgeris2020,wu2020mag,testbed2,
Basar2019access,Ref3a00,Ref3a01,Ref3a,Ref4a,Ref5,Ref6,yang2020ee}. Specifically, the reflecting RIS usually is implemented as passive reflecting arrays to assist transmissions. A typical RIS that likes an intelligent mirror usually consists of a large number of nearly passive, low-cost, and low-energy consuming reflecting elements, each of which can change the direction of EM signals impinging on it \cite{chongwenhmimos,wu2020mag,meta1,meta2}. By changing the certain phase shifts, the reflected signals can be focused  constructively to the desired user for enhancing the received signal power or destructively at non-intended users to reduce the co-channel interference\cite{chongwenhmimos,georgeris2020,liaskos2018,Ref9,wu2020mag}. Benefited from its some features, the passive RIS can be fabricated in very compact size with light weight, resulting in 1) easy installation in building facades, ceilings of training, walls, laptops, clothes, etc., 2) ready integration into existing communication systems without the modifications on hardware, as well as 3) compact deployment of multiple RISs along the signal propagation path\cite{chongwenhmimos,wu2020mag,meta1,meta2,Ref8a,Ref9,Ref10,Basar2019access,Ref10a,testbed2,Ref10b,Ref10c,Ref10d,Ref10e,Ref10f,Ref10g}.

\subsection{Prior Works}
RISs have aroused widely public concern and interesting recently, but most of the reported works mainly focus on the optimal RIS phase shifts and  beamforming design to improve the performance of wireless communication \cite{Ref5,Ref6}, \cite{Ref9,Ref10a,Ref10b,smith2020tsp}. For example, the potential transmission and positioning for RIS assisted multiple input multiple output (MIMO) systems  were investigated in \cite{Ref5}, where the RIS is used as a transmitter. \cite{Ref9} proposed an index modulation (IM) scheme by leveraging the programmable inherit characteristic of RISs. Simulation results show that the RIS-based IM can realize  high data rates with significantly low bit error rates. By taking full advantage of its low-power consumption, our previous work \cite{chongwentwc2019} also shows that RIS-based wireless systems can improve the energy efficiency triple times more than the traditional relay-based systems. To maximize the sum rate of considered RIS assisted single user multiple input single output (MISO) systems, \cite{Ref10a} proposed an alternating optimization  method to find the local optimal transmit beamforming at the base station (BS) and phase shifts design of RISs.

Recently, RIS-based designs have emerged as strong candidates that empower communications at the THz band. Specifically, \cite{Akyildiz_Thz2018wcm,Sarieddeen2020wcm,han2019terahertz,sarieddeen2020overview} presented some promising visions and potential applications leveraging the advances of RIS to combat the propagation attenuations and molecular absorptions of THz frequencies. To remove obstacles of realizing these applications, \cite{cemaxinying2020,ning2019channel,wei2020ce,wei2020cetcom,xiaojun2020ce} proposed the channel estimation and data rate
maximization transmission solutions for massive MIMO RIS-assisted THz system. Furthermore, some beamforming and resource allocation schemes were proposed in \cite{ning2019beamforming,Nie2020beamforming,Nie2019resource}. For example,  a cooperative beam training scheme and two cost-efficient hybrid beamforming schemes were proposed in \cite{ning2019beamforming} for the THz multi-user massive MIMO  system with RIS, while a resource allocation based on the proposed end-to-end physical model was introduced in \cite{Nie2019resource} to  improve the achievable distance
and data-rate at THz band RIS-assisted communications. What's more, a RIS-based scheme to improve the coverage range for the indoor THz communication scenarios was reported in \cite{ma2020intelligent}, and \cite{qiao2020secure} revealed that RIS-based secure strategy can significantly improve the secrecy performance of THz communication systems.

All above works assume single-hop RIS assisted systems, where only one RIS is deployed between the BS and the users. In practical, similar to multi-hop relaying systems, multiple RISs can be used to overcome severe signal blockage between the BS and users to achieve better service coverage. Although multi-hop MIMO relaying systems have been addressed in the literature intensively in the context of relay selection, relay deployment, and precoding design, multi-hop RIS empowered systems have rarely studied. In multi-hop relay systems, with ready installation of RF chains, the signal arrived at the relays can be processed and optimally designed prior to be retransmitted to the users, given certain predetermined constraints. Various active beamforming techniques can be fully exploited to achieve better performance. However, the methodologies developed for multi-hop relay systems cannot be directly applied to multi-hop RIS assisted systems, due to different reflecting mechanisms and channel models. Particularly, the constraint on diagonal phase shift matrix and unit modulus of the reflecting RIS makes the joint design of transmit beamforming and phase shifts extremely challenging. To the authors' best knowledge, multi-hop multiuser RIS assisted MISO systems have not yet been addressed in the existing literature.

To address high-dimension, complex EM environment, and mathematically intractable non-linear issues of communication systems, the model-free machine learning method as an extraordinarily remarkable technology has introduced in recent years \cite{Ref14,Ref14a,mnih2015human,Ref14c,deeplearn2,chen2019FLwireless}.  Overwhelming research interests and results uncovers  machine learning technology to be used in the future 6G wireless communication systems for dealing with the non-trivial problems due to extremely large dimension in large scale MIMO systems \cite{Ref15,Ref16g}. To be specific, deep learning has been used to obtain the channel state information (CSI) or beamforming matrix in non-linear communication systems \cite{Ref16a,Ref16b,Ref16c,Ref16d}. In terms of dynamic and mobile wireless scenarios, deep reinforcement learning (DRL) provides an effective solution by leveraging the advantages of deep learning, iterative updating and interacting with environments over the time \cite{Ref15a,Ref15,Ref14c,deeplearn2,nie2019dl,liaskos2019dnn,Ref16a,Ref16b,Ref16c,Ref16d,Ref16e,Ref16f,Ref16g,chenxianfu_drl}. In particular, the hybrid beamforming matrices were obtained by DRL for the mobile mmWave systems in \cite{Ref16e}, while   \cite{Ref16g} proposed a novel idea to utilize DRL for optimizing the network coverage. What's more, \cite{Ref16f} used the DRL to tackle the joint non-convex optimization problem by considering complex constraints of resource allocation, beamforming interference coordination, etc.

\subsection{Contributions}

In this paper, we present a multi-hop RIS-assisted communication scheme to overcome the severe propagation attenuations and improve the coverage range at THz-band frequencies, where the hybrid design of transmit beamforming at the BS and phase shift matrices is obtained by the advances of DRL. Specifically, benefited from the recent breakthrough on RIS, our main objective is to overcome propagation attenuations at THz-band communications by deploying multiple passive RISs  between the BS and multiple users. To maximize the sum rate, formulated optimization problem is non-convex due to the multiuser interference, mathematically intractable multi-hop signals, and non-linear constraints. Owning to the presence of possible multi-hop propagation, which results in composite channel fadings, the optimal solution is unknown in general. To tackle this intractable issue, a DRL-based algorithm is proposed to find the feasible solutions. The main contributions of this paper are given as follows:

$\bullet$ We propose a practical RIS-empowered hybrid beamforming architecture improve the coverage range for THz-band communications. Based on the framework, we formulate a non-convex joint design problem of the digital beamforming and analog beamforming matrices.

$\bullet$ To tackle this NP-hard problem, a DRL-based algorithm is proposed, which is a very early attempt to address the joint design of the digital beamforming and analog beamforming matrices for multi-hop RIS assisted THz-band communication systems.

$\bullet$ To improve the convergence and overcome the local optimal solution of the proposed DRL algorithm, two methods are proposed to initialize the digital beamforming and analog beamforming matrices.

$\bullet$ Simulation results show that our proposed  algorithms not only are able to overcome the propagation attenuations and improve the 50\% more coverage range of THz communications, but also are a state-of-the-art method to solve the NP-hard beamforming problem, especially when the signals at RIS-empowered THz communication networks experience multiple hops. 

The outline of this work is given as follows. The THz system model and problem formulation is introduced in Section II. Before the DRL-based design of digital and analog beamforming is proposed in Section IV, the framework of the proposed DRL and policy is presented in Section III. Simulation results are presented in Section V to verify the performance of the proposed scheme, whereas conclusions are drawn in Section VI.

The notations of this paper are summarized as follows. We use the $\mathbf{H}$ to denote a general matrix, and $\mathbf{H}(i,j)$ denotes the entry at the $i^{th}$ row and the $j^{th}$ column. $\mathbf{H}^{(t)}$ is the value of $\mathbf{H}$ at time $t$. $\mathbf{h}_k$ is the $k^{th}$ column vector of $\mathbf{H}$. $\mathbf{H}^T$, and $\mathbf{H}^{\mathcal{H}}$ denote the transpose and conjugate transpose of matrix $\mathbf{H}$, respectively. $Tr \{ \}$ is the trace of the enclosed. For any vector $\mathbf{g}$, $\mathbf{g}(i)$ is the $i^{th}$ entry, while $\mathbf{g}_{k}$ is the channel vector for the $k^{th}$ user. $||\mathbf{h}||$ denotes the magnitude of the vector. $\mathcal{E}$ denotes statistical expectation. $|x|$ denotes the absolute value of a complex number ${x}$, and its real part and imaginary part are denoted by $Re{(x)}$ and $Im{(x)}$, respectively.

\vspace{-0.15cm}
\section{System Model and Problem Formulation}

\subsection{Terahertz-Band Channel Model}

Unlike the lower frequency band communications, a signal operating at the THz band can be affected easily by many peculiarly factors, mainly is influenced by the molecular absorption due to water vapor and oxygen, which result in very high path loss for line-of-sight (LoS) links\cite{rappaport2019,han2018mag,han2016dis,han2014multi,Moldovan2016,Moldovan2017,Nie2020beamforming}. On the other hand, spreading loss also contributes a large proportion of attenuations. In terms of  non-line-of-sight (NLoS) links, besides mentioned peculiarities, unfavorable material and roughness of the reflecting surface also will cause a very severe reflection loss\cite{Moldovan2016,Moldovan2017,han2014multi,Nie2020beamforming}. The overall channel transfer function can be written as,

\begin{equation} \label{eq:modelthz1}
\begin{split}
H(f,d,\bm{\zeta})=H^{LOS}(f,d)e^{-j2\pi f\tau_{LOS}}+\sum^{M_{rays}}_{i=1}H_i^{NLOS}(f,\zeta_i)e^{-j2\pi f\tau_{NLOS_i}},
\end{split}
\end{equation}
where $f$ denotes the operating frequency, $d$ is the distance between the transmitter and receiver, the vector $\bm{\zeta}=[\zeta_1,...,\zeta_{M_{rays}}]$ represents the coordinates of all scattering points, and $\tau_{LOS}$ and $\tau_{NLOS_i}$ denote  the propagation delays of the LOS path and $i^{th}$ NLOS path respectively.

The frequency response for the LOS channel $H^{LOS}(f,d)$ is given as\cite{Moldovan2016,han2014multi},
\begin{equation} \label{eq:modelthz2}
H^{LOS}(f,d)=H_{spread}(f,d)\cdot H_{molec}(f,d),
\end{equation}
with
\begin{align} \label{eq:modelthz3}
H_{spread}(f,d)&=\frac{c}{4\pi \cdot f \cdot d}, \\
H_{molec}(f,d) &=e^{-\frac{1}{2}\alpha_{molec}(f,T_k,p)d},
\end{align}
where $c$ is the light speed, $\alpha_{molec}(f,T_k,p)$ is the total molecular absorption loss with gas pressure $p$ and temperature $T_k$. Similarly, the frequency response for the $i^{th}$ NLOS channel $H_i^{NLOS}(f,\zeta_i)$ is given by\cite{Moldovan2016,han2014multi},
\begin{equation} \label{eq:modelthz4}
H_i^{NLOS}(f,\zeta_i)=H_{relf,i}(f,d_{i2},\vartheta_{i1},\vartheta_{i2},\vartheta_{i3})\cdot H_{spread,i}(f,d_{i1},d_{i2})\cdot H_{molec,i}(f,d_{i1},d_{i2}),
\end{equation}
with
\begin{align} \label{eq:modelthz5}
H_{relf,i}(f,d_{i2},\vartheta_{i1},\vartheta_{i2},\vartheta_{i3})&=
\sqrt{\mathcal{E}\{R_{power,i}(f,d_{i2},\vartheta_{i1},\vartheta_{i2},\vartheta_{i3})\}}\\
H_{spread,i}(f,d_{i1},d_{i2})&=\frac{c}{4\pi \cdot f \cdot (d_{i1}+d_{i2})},\\
H_{molec,i}(f,d_{i1},d_{i2})&=e^{-\frac{1}{2}\alpha_{molec}(f,T_k,p)(d_{i1}+d_{i2})},
\end{align}
where $\mathcal{E}\{R_{power,i}(f,d_{i2})\}$ denotes the scattered power on a surface area.

\subsection{Proposed Multi-hop Scheme}

\begin{figure}[ht]
\begin{center}
  \includegraphics[width=17cm]{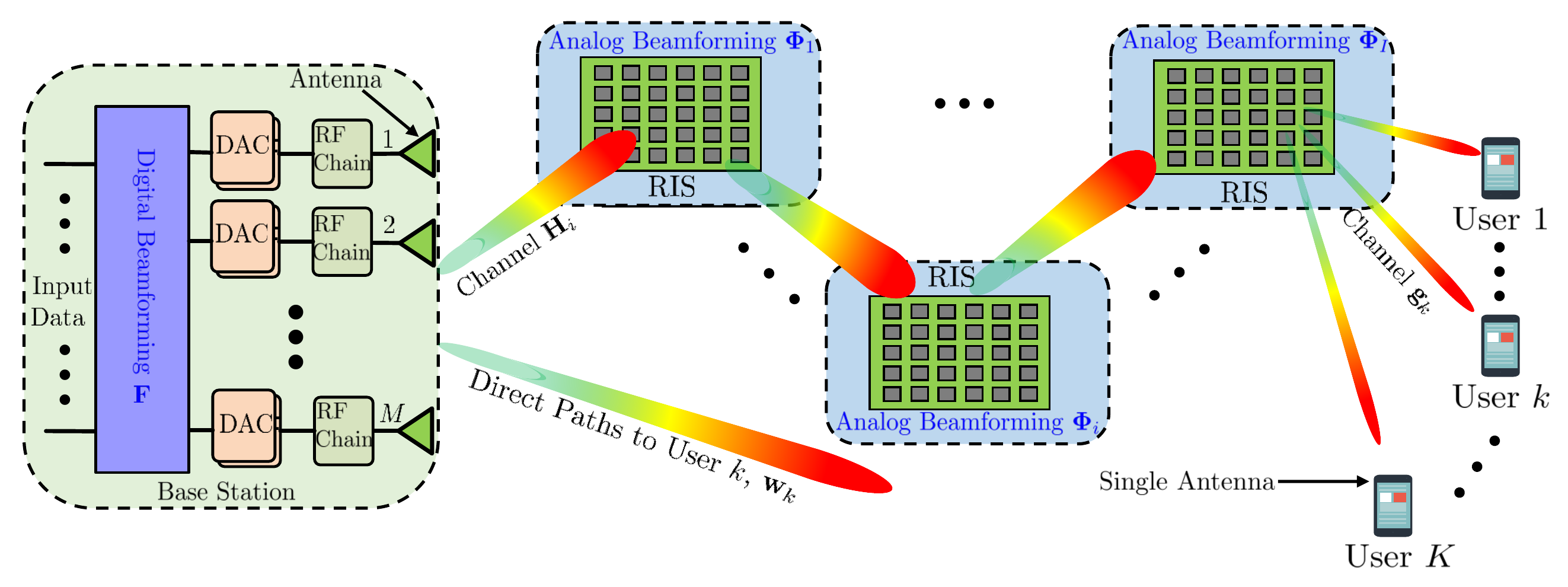}
  \caption{The RIS-based multi-hop for THz communications and proposed practical RIS-based hybrid beamforming architecture. In contrast to the traditional beamforming architectures,  a key novelty of this scheme is to take full advantage of RISs with the unique programmable feature as an external and standalone analog beamforming, which not only can remove internal analog beamforming at BS that simplifies the architecture and reduces cost significantly, but also improve the beamforming performance of THz-band communication systems significantly.   }
  \label{fig:hybrid}
\end{center}  \vspace{-6mm}
\end{figure}

As mentioned before, communications over the THz band are very different with the low frequency band communications, the transmitted signal suffers from the severe path attenuations\cite{rappaport2019,han2018mag,han2016dis}. To address this issue, we introduce a multi-hop multiuser system by leveraging some unique features of RISs, which is comprised of a BS, $N$ reflecting RISs and multiple single-antenna users shown in Fig. \ref{fig:hybrid}. We consider that BS equipped with $M$ antennas communicate with $K$ single-antenna users in a circular region. Assume that the $i^{th}$ reflecting RIS, $i=1,\cdots, N$, has $N_i$ reflecting elements. A number of $K$ data streams are transmitted simultaneously from the $M$ antennas of the BS with the aid of multiple RISs to improve the coverage range of THz communications. Each data stream is beamforming to one of the $K$ users by the assistance of RISs.  In other words, signals are first arrived at the reflecting RIS and then reflected by this RISs. In order to enhance the transmission distance and overcome the severe path loss, multi-RISs are deployed randomly with the programable feature to collect more energy for the object user.


Although we have this multi-hop scheme, how we implement it from the practical hardware perspective and perform the effective beamforming are challenging, since traditional hybrid precoding schemes are too expensive for multi-hop THz communications. Besides the multi-hop scheme, a practical RIS-empowered hybrid precoding architecture is also shown in Fig. \ref{fig:hybrid}. The direct transmission paths between the BS and users are also considered here although they are usually week after the long distanced transmissions. In contrast to the traditional precoding architectures,  a key novelty of this scheme is to take full advantages of RISs with the unique programmable feature as an external and portable analog precoder, i.e., the RIS functions as a reflecting array, equivalent to introduce the analog beamfroming to impinge signals, which not only can remove internal analog precoder at BS that simplifies the architecture and reduces cost significantly, but also improve the beamforming performance of THz-band communication systems.


We assume that the channel fading is frequency flat, and the transmitted signal experiences $I_k (I_k \leq N)$ hops on RISs to arrive $k^{th}$ user. We denote the channel matrix from the BS to the first reflecting RIS as $\mathbf{H}_1 \in \mathbb{C}^{(N_1 \times M )}$, the channel matrix from the $i^{th}$ RIS to the $(i+1)^{th}$ RIS as $\mathbf{H}_{(i+1)} \in \mathbb{C}^{(N_{(i+1)} \times N_i)}$.  The received signal at the $k^{th}$ user is given as
\begin{equation} \label{eq:sys_1}
\begin{split}
y_k=(\mathbf{g}_{k}^T \prod_{i=1,\cdots, I_k} \mathbf{\Phi}_i \mathbf{H}_{i}+\mathbf{w}_k)\mathbf{x}+n_k \\
\end{split}
\end{equation}
where the vector $\mathbf{g}_{k} \in \mathbb{C}^{(N_{I_k} \times 1)}$ and $\mathbf{w}_k \in \mathbb{C}^{( 1 \times M )} $ denote the channel from the last RIS to the $k^{th}$ user and the direct channel from the BS to user $k$ respectively, $\mathbf{\Phi}_i \triangleq\mathrm{diag}[\theta_{i1},\theta_{i1},\ldots,\theta_{iN_i}] \in \mathbb{C}^{(N_{i} \times N_{i})}$ is the phase shift matrix of the $i^{th}$ RIS, i.e., the $i^{th}$ analog precoding matrix, $\mathbf{x} \in \mathbb{C}^{M \times 1} $ is the transmit vector from the BS, and $n_k$ is the additive white Gaussian noise (AWGN) with the zero mean and $\sigma_n^2$ variance. For the higher communications, like THz, the channels are perhaps dominated by the line-of-sight (LOS) paths. To match the practical implementation, we employ the Rician fading to model the channels $\mathbf{H}_{i}$, which can be written as
\begin{equation} \label{eq:sys_11}
\begin{split}
\mathbf{H}_{i}=\sqrt{\frac{K_{H}}{K_{H}+1}}\mathbf{\overline{H}}_{i}+\sqrt{\frac{1}{K_{H}+1}}\mathbf{\widetilde{H}}_{i},
\end{split}
\end{equation}
where $K_{H}$ is the Rician factor of $\mathbf{H}_{i}$; $\mathbf{\overline{H}}_{i}$ is LoS component, which remains unchanged within the channel coherence time, and $\mathbf{\widetilde{H}}_{i}$ is the non-LoS (NLoS) component. The elements of $\mathbf{\widetilde{H}}_{i}$ are modeled as the complex Gaussian distributed with the zero mean and unit variance. Similarly, the channels $\mathbf{g}_{k}$ and $\mathbf{w}_k$ are also modeled as the Rician distribution with the Rician factor as $K_{g}$ and $K_{w}$ respectively.

We further assume that the channel $\mathbf{g}_{k}$, $\mathbf{w}_{k}$, and $\mathbf{H}_{i}$ for all $K$ users are perfectly known at both the BS and all users. Although we admit that obtaining these CSIs are  challenging tasks for RIS-based communication systems, there are already significant methods that are proposed in existing works. For example, we proposed an alternating least square method based on the parallel factor framework in \cite{wei2020ce,wei2020cetcom}, which continuously estimates the all channels without too high complexity. Furthermore, research on the channel estimation is also beyond the scope of this paper. Therefore, we have this assumption. The transmit vector $\mathbf{x}$ can be written as:
\begin{equation} \label{eq:sys2}
\mathbf{x}\triangleq\sum_{k=1}^{K}\mathbf{f}_{k}s_{k}
\end{equation}
where $\mathbf{f}_{k}\in\mathbb{C}^{M\times 1}$ and $s_{k} \in \mathcal{CN}(0,1)$, i.e., under the assumption of Gaussian signals, denote the beamforming vector and independent user symbols respectively.  The power of the transmit signal from the BS has the following constraint:
\begin{equation}\label{eq:sys3}
  \mathcal{E}[|\mathbf{x}|^2]=\mathrm{tr}(\mathbf{F}^H\mathbf{F})\leq P_t\;,
\end{equation}
wherein $\mathbf{F}\triangleq[\mathbf{f}_1,\mathbf{f}_2,...,\mathbf{f}_K]\in\mathbb{C}^{M\times K}$, and $P_t$ is the total transmission power of the BS.

It should be noted that $\mathbf{\Phi}_i$ is a diagonal matrix whose entries are given by $\mathbf{\Phi}_i(n_i,n_i)=\theta_{in_i}=e^{j\phi_{n_i}}$, where $\phi_{n_i}$ is the phase shift induced by each element of the RIS. Like a mirror, the signal goes through the RIS is no energy loss, which means  $|\mathbf{\Phi}_i(n_i,n_i)|^2$=1. In this paper, we consider each RIS that has the infinite  phase shift resolution, i.e., $\phi_{n_i} \in [0,2\pi) \forall n$ for the development of DRL-based algorithm. It also can be seen that, the signals arrive at the object users experience the composite channel interference, $\sum_{j,j\neq k}^K\mathbf{g}_{k}^T \prod_{i=1,\cdots, I_k} \mathbf{\Phi}_i \mathbf{H}_{i} \mathbf{f}_jx_j$. 
The received signal (\ref{eq:sys_1}) can be further given as
\begin{equation} \label{eq:sys_1a}
\begin{split}
y_k=\bigg(\mathbf{g}_{k}^T \prod_{i=1,\cdots, I_k} \mathbf{\Phi}_i \mathbf{H}_{i}+\mathbf{w}_k \bigg) \mathbf{f}_kx_k+\sum_{j, j\neq k}^K \bigg(\mathbf{g}_{k}^T \prod_{i=1,\cdots, I_k} \mathbf{\Phi}_i \mathbf{H}_{i}+\mathbf{w}_k \bigg) \mathbf{f}_jx_j+n_k \\
\end{split}
\end{equation}
where $\mathbf{f}_m$ is the beamforming vector for the $m^{th}, m\neq k$ user.
Furthermore, the SINR at the $k^{th}$ user is written as
\begin{equation} \label{eq:sys_3}
\rho_{k}=\frac{|(\mathbf{g}_{k}^T \prod_{i=1,\cdots, I_k} \mathbf{\Phi}_i \mathbf{H}_{i}+\mathbf{w}_k) \mathbf{f}_k|^2}{|(\sum_{j,j\neq k}^K\mathbf{g}_{k}^T \prod_{i=1,\cdots, I_k} \mathbf{\Phi}_i \mathbf{H}_{i}+\mathbf{w}_k) \mathbf{f}_j|^2+\sigma_n^2}
\end{equation}

\subsection{Problem Formulation}

Our main objective is to combat the propagation attenuations of THz communications by leveraging multi-hop RIS-assisted communication scheme. Therefore, we use the ergodic sum rate as evaluate metric, and we have
\begin{align} \label{eq:sys_6}
C(\mathbf{F},\mathbf{\Phi}_{i,\forall i}, \mathbf{w}_{k, \forall k}, \mathbf{g}_{k, \forall k},\mathbf{H}_{i,\forall i}) &=\sum_{k=1}^K R_k \\
&=\sum_{k=1}^K\log_2(1+\rho_k)
\end{align}
where we use $R_k$ to denote the data rate at the $k^{th}$ user. It is clear to see that the major obstacles to maximize the sum rate are to give the optimal design of digital beamforming matrix $\mathbf{F}$ and beamforming matrix $\mathbf{\Phi}_{i}, \forall i$, i.e., phase shift matrix of RISs. Therefore, the main problems for reducing the propagation attenuation of multi-hop RIS-assisted THz communication system are changed to obtain the optimal $\mathbf{F}$ and $\mathbf{\Phi}_{i}, \forall i$. Then, we have the following optimization problem as,
\begin{equation} \label{eq:BD_1}
\begin{split}
&  \max\limits_{\mathbf{F},\mathbf{\Phi}_i} C(\mathbf{F},\mathbf{\Phi}_{i,\forall i}, \mathbf{w}_{k, \forall k}, \mathbf{g}_{k, \forall k},\mathbf{H}_{i,\forall i})  \\
& \; \textrm{s.t.} \;\;  tr\{\mathbf{F}\mathbf{F}^{\mathcal{H}} \} \leq P_t \\
& \;\;\;\quad\;\; |\phi_{in_i}|=1\;\forall n_i=1,2,\ldots,N_i,\\
\end{split}
\end{equation}
Unfortunately, we can easily find that the optimization problem (\ref{eq:BD_1}) is a NP-hard problem because of the non-trivial objective function and the non-convex constraint. As we all know, it is nearly impossible to obtain an analytical solution by the traditional methods of mathematical analysis for the  multi-hop optimization. In addition, exhaustive numerical search is also impractical for large scale networks. Although there are some existing approximation methods that are proposed based on the alternating method to find the sub-optimal solutions for single hop RIS-based system, e.g., \cite{Ref10a,Ref10b,Ref10c,Ref10d,Ref10e,Ref10f,Ref10g}, they are difficult to work for the multi-hop scenario, especially we do not know how many RIS hops the transmitted signal experienced  to arrive $k^{th}$ user, i.e., $I_k (I_k \leq N)$ in prior. Instead, in this paper, we will propose a new method by leveraging the recent advance on DRL technique, rather than directly solving this challenging optimization problem mathematically. Unlike the traditional deep neural network (DNN), the proposed DRL does not need two phases, i.e., offline training and online learning, but can obtain the these matrices continuously by giving the CSI. The details of the proposed DRL-based method will be given in the following.

\section {Framework of the Proposed DRL and Policy}
In this section, we introduce the framework of the proposed DRL and optimization policy, which is utilized to design the transmit beamforming matrix and phase shift matrices in multi-hop THz communication networks.

\subsection {Framework of DRL}


One of key benefits in developing artificial intelligence is its ability to solve intricate optimization problems with high-dimensional input and output spaces. Recently, many significant progress has been done on reinforcement learning (RL) by leveraging the advance of deep learning, resulting in the DRL that is applied in the wireless communication field with a variety of optimization goals. Generally, a typical DRL framework consists of six fundamental elements,  i.e., the state set $\mathbf{S}$, the action set $\mathbf{A}$, the instant reward $r(s,a), (s\in \mathbf{S}, a \in \mathbf{A})$, the policy $\pi(s,a)$, transition function $\mathbf{P}$ and Q-function $Q(s,a)$.  Note that the policy $\pi(s,a)$ denotes the conditional probability of taking action $a$ on the instant state $s$. This also means that the policy $\pi(s,a)$  needs to satisfy $\sum_{a \in \mathbf{A}, s \in \mathbf{S},}\pi(s,a) =1$. In addition, since we consider a mobile environment, the transition function $\mathbf{P}$ usually is affected by the environment itself and the action form the RL agent, which can be modeled by the Markov decision process (MDP).

We adopt the maximization of the average rewards as the action policy of RL agent, since the average rewards is not only affected by the instant rewards but also the future rewards, which can avoid the instant interference. The future cumulative reward can be given at time $t$ as
\begin{equation} \label{eq:drl1}
\begin{split}
 R^{(t)}=\sum_{\tau=0}^\infty \beta^{\tau} r^{t+\tau+1}
\end{split}
\end{equation}
where $\beta$ is a weighting coefficient for the future rewards. The typical Q-function paired with the policy $\pi(s,a)$ is defined as $Q_{\pi}(s^{(t)},a^{(t)})$. Then, once the action $a$ is taken at state $s$, its Q-function is written as
\begin{equation} \label{eq:BD_31}
\begin{split}
Q_{\pi}(s^{(t)},a^{(t)})=\mathcal{E}_{\pi}\big [ R^{(t)}|s^{(t)}=s, a^{(t)}=a \big ]
\end{split}
\end{equation}

Introducing the action-state transition function $P_{ss'}^a=\textrm{P}r(s^{(t+1)}=s'|s^{(t)}=s, a^{(t)}=a)$ from the state $s'$ to $s$, then the Q-function can further is given as,
\begin{equation} \label{eq:BD_31a}
\begin{split}
Q_{\pi}(s^{(t)},a^{(t)})=&\mathcal{E}_{\pi}\big [ r^{(t+1)} |s^{(t)}=s, a^{(t)}=a \big ]+\\
&\beta \sum_{s' \in \mathbf{S}}P_{ss'}^a \bigg (\sum_{a' \in \mathbf{A}} \pi(s', a')Q^{\pi}(s',a') \bigg)
\end{split}
\end{equation}

Under the objective of maximizing the expectation of the instant rewards for the environmental state $s$, the RL agent searches the optimal policy $\pi^*$. Therefore, the optimal $Q^*(s^{(t)},a^{(t)})$ function will becomes as
\begin{equation} \label{eq:BD_31b}
\begin{split}
Q^*(s^{(t)},a^{(t)})=&r^{(t+1)}(s^{(t)}=s, a^{(t)},\pi=\pi^*) +\\
&\beta \sum_{s' \in \mathbf{S}}P_{ss'}^a \max_{a' \in \mathbf{A}} Q^*(s',a')
\end{split}
\end{equation}

\begin{figure}
\begin{center}
  \includegraphics[width=17cm]{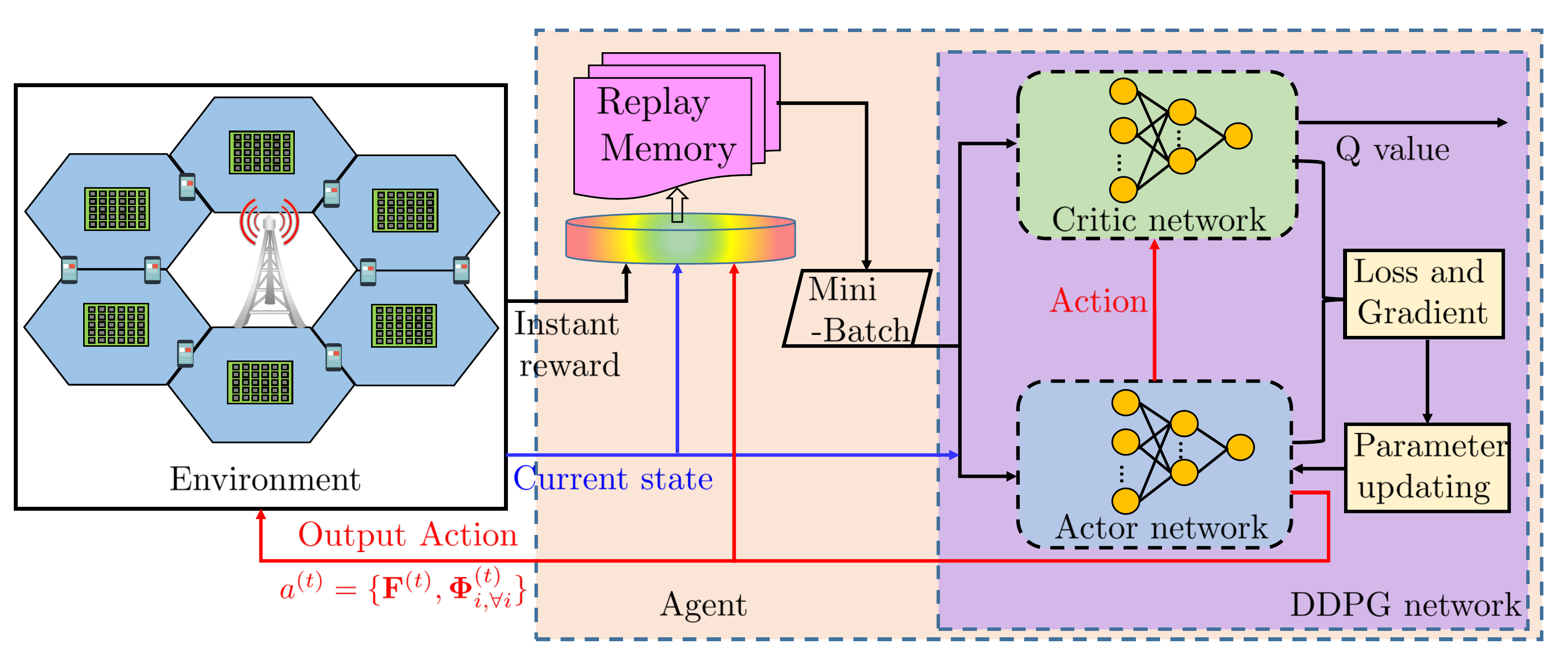}
  \caption{ The illustration of the proposed DRL framework, and proposed actor-critic DDPG algorithm. }
  \label{fig:NN} \vspace{-8mm}
\end{center}
\end{figure}
However, it is very challenging to directly find the optimal $Q^*(s^{(t)},a^{(t)})$ due to the unknown $P_{ss'}^a $ action-state transition function. The classical Q-learning is to construct a Q-table, which is a discrete set of $|\mathbf{S} \times \mathbf{A}| $ with the random initialization. Adopting an $\epsilon-$greedy policy to take the action according to each environmental change, the updating for Q-table is listed as
\begin{equation} \label{eq:BD_32a}
\begin{split}
Q^{*}(s^{(t)},a^{(t)})\leftarrow &(1-\alpha)Q^{*} (s^{(t)},a^{(t)})+ \alpha (r^{(t+1)}+\\
&\gamma \max_{a'}Q_{\pi}(s^{(t+1)},a'))
\end{split}
\end{equation}
where $\alpha \in (0,1]$ denotes the learning rate. Regarding to our proposed hybrid beamforming problem that have an approaching an infinite state and action space, the storage size and search complexity of Q-table are extremely impractical. To overcome these issues, we employ a deep Q-learning method to approximate the Q-table by leveraging the universal approximation feature of DNNs \cite{universal}. As shown in Fig. \ref{fig:NN}, our proposed DRL framework uses two DNNs (also named actor network and critic network)  to approximate the state/action value function. In other words, a actor neural network to approximate a policy based on the observed environment $s$ state and output an action, while another DNN implements the critic network denoted $Q(\mathbf{\theta} |s(t),a(t))$ to evaluate  the current policy according to the received the rewards. Specifically, the critic network is to update value function approximation, and its  policy parameters are updated by the actor network. The weight parameter $\mathbf{\theta}$ is used to construct the policy $\pi(\theta|s, a)$, and it is updated by the following gradient rule,
\begin{equation} \label{eq:BD_32c}
\begin{split}
\mathbf{\theta}^{(t+1)}=\mathbf{\theta}^{(t)}-\mu \Delta_{\mathbf{\theta}} \ell (\mathbf{\theta})
\end{split}
\end{equation}
where $\mu$ and $\Delta_{\mathbf{\theta}}$ denote the learning rate and gradient of loss function $\ell (\mathbf{\theta})$ respectively. Since we do not know the optimal policy and the optimal target during the reinforcement learning, we definite the two identical networks as target Q-function $Q(\mathbf{\theta}^{(train)}|s^{(t)},a^{(t)})$ and training Q-function $Q(\mathbf{\theta}^{(target)}|s^{(t)},a^{(t)})$, and the differences between them are defined as the loss function. Assume that the training Q-function network is synchronized with  the target Q-function network. And the loss function can be written as
\begin{equation} \label{eq:BD_32c2}
\begin{split}
&\ell (\mathbf{\theta})=\bigg ( y- Q(\mathbf{\theta}^{(train)}|s^{(t)},a^{(t)}) \bigg )^2
\end{split}
\end{equation}
where the target function $y$ are updated by
\begin{equation} \label{eq:BD_32c1}
\begin{split}
y=\left( r^{(t+1)}+\beta \max_{a'} Q(\mathbf{\theta}^{(target)}|s^{(t+1)},a'\right)
\end{split}
\end{equation}

\subsection{Optimization Policy}
As shown in Fig. \ref{fig:NN}, our proposed hybrid beamforming design problem can be resorted the recent advance of DRL to interact the environment, but one of key problems is that the action and state spaces are not the prior knowledge, and also continuous. In addition, it is well known that the traditional DQN just be good for the problems with a few discrete state spaces, but not for continuous case. Fortunately, the recent proposed method named DDPG or deep deterministic policy gradient derived from the MDP can solve this issue \cite{casas2017deep,gu2016q}. Furthermore, it also can combine the actor-critic reinforcement learning framework. Based on this, we can update the training critic network based on the DDPG policy as follows:
\begin{equation} \label{eq:BD_32d}
\begin{split}
&\mathbf{\theta}_c^{(t+1)}=\mathbf{\theta}_c^{(t)}-\mu_c \Delta_{\mathbf{\theta}_c^{(train)}} \ell (\mathbf{\theta}_c^{(train)})
\end{split}
\end{equation}
\begin{equation} \label{eq:BD_32d1}
\begin{split}
\ell (\mathbf{\theta}_c^{(train)})=&\bigg ( r^{(t)}+\beta q(\mathbf{\theta}_c^{(target)}|s^{(t+1)},a')-\\ &q(\mathbf{\theta}_c^{(train)}|s^{(t)},a^{(t)}) \bigg )^2
\end{split}
\end{equation}
where $\mu_c$  and  $a'$ denote the learning rate and  the action coming from the target actor network respectively, and $\Delta_{\mathbf{\theta}_c^{(train)}} \ell (\mathbf{\theta}_c^{(train)})$ represents the gradient of training network. The $\mathbf{\theta}_c^{(target)}$ and the $\mathbf{\theta}_c^{(train)}$ represent the policy parameters of training and the target critic networks respectively.

\begin{algorithm}[H]
\caption{DDPG algorithm}
\label{alg:ALGddpg}
\textbf{Initialization:} Given initialized $Q(\mathbf{\theta}^{(train)}|s^{(t)},a^{(t)})$  and $Q(\mathbf{\theta}^{(target)}|s^{(t)},a^{(t)})$ \\
Initialize the $\mathbf{\theta}_a^{(target)}$  and $\mathbf{\theta}_a^{(train)}$
\begin{algorithmic}[1]
\FOR{ \texttt{episode $=0,1,2, \cdots, Z-1$} }
\STATE Collect initial state $s$ and action $a$ \\
\FOR{\texttt{t=$0,1,2, \cdots, T-1$}}
\STATE Select action $a^{(t+1)}$ = $\pi(\theta|s^t, a^t)$ from the actor network \\
\STATE Observe new state $s^{(t+1)}$ and rewards $r^{(t+1)}$ \\
\STATE Save the experience $(s^{(t)}, a^{(t)}, r^{(t+1)}, s^{(t+1)})$ in the memory\\
\STATE Sample from the minibatch\\
\STATE Set target function by the equation (\ref{eq:BD_32c1}) \\
\STATE Minimizing the loss function at critic network by the equation (\ref{eq:BD_32d1})\\
\STATE Update actor networks based policy gradient by the equation (\ref{eq:BD_32e})\\
\STATE Update target networks by the equation (\ref{eq:BD_32f}) \\
\ENDFOR
\ENDFOR
\end{algorithmic}
\end{algorithm} \vspace{-4mm}
We assume to run the algorithm over $Z$ episodes, and each episode involves the $T$ iterations. Similarly, we can update the training actor network based on the DDPG policy as Algorithm \ref{alg:ALGddpg}.
\begin{equation} \label{eq:BD_32e}
\begin{split}
&\mathbf{\theta}_a^{(t+1)}=\mathbf{\theta}_a^{(t)}-\mu_a \Delta_{a}q(\mathbf{\theta}_c^{(target)}|s^{(t)},a) \Delta_{\mathbf{\theta}_a^{(train)}}\pi (\mathbf{\theta}_a^{(train)}|s^{(t)})
\end{split}
\end{equation}
where $\mu_a$ denotes the learning rate, and  $\pi (\mathbf{\theta}_a^{(train)}|s^{(t)})$ denotes the training actor parameters at the state $s$. The gradient of target critic network is denoted as $\Delta_{a}q(\mathbf{\theta}_c^{(target)}|s^{(t)},a)$, while $\Delta_{\mathbf{\theta}_a^{(train)}}\pi (\mathbf{\theta}_a^{(train)}|s^{(t)})$ denotes the gradient of training actor network under the parameter of $\mathbf{\theta}_a^{(train)}$. We can seen from above the equation, the gradients of target critic network and training actor network  both have impact on the updating of training actor network, which also make sure that the actor-critic networks is toward right direction to converge for learning the continuous-valued spaces.

Finally, the target actor and critic network can be updated by the following equations,
\begin{equation} \label{eq:BD_32f}
\begin{split}
& \mathbf{\theta}_a^{(target)}\leftarrow \tau_a \mathbf{\theta}_a^{(train)}+(1-\tau_a)\mathbf{\theta}_a^{(target)}  \\
& \mathbf{\theta}_c^{(target)}\leftarrow \tau_c \mathbf{\theta}_c^{(train)}+(1-\tau_c)\mathbf{\theta}_c^{(target)} \\
\end{split}
\end{equation}
where the learning rate of target actor and critic network are denoted by $\tau_a$ and $ \tau_c$. The DDPG algorithm are summarized as Algorithm \ref{alg:ALGddpg}.

\section {DRL-based Design of Digital and Analog Beamforming}
\begin{figure}
\begin{center}
  \includegraphics[width=12.5cm]{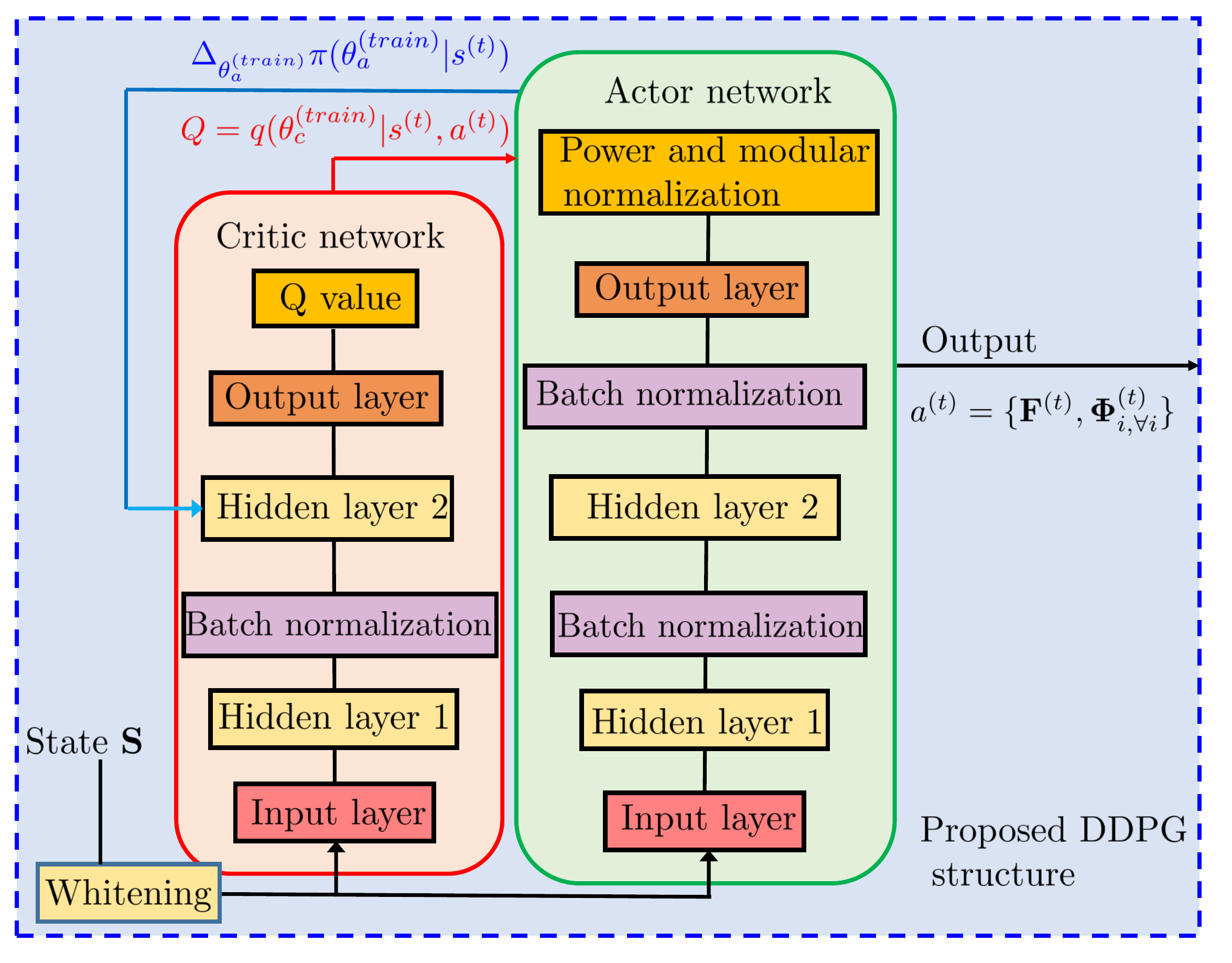}
  \caption{ Proposed DDPG structure including the critic neural network
and the actor neural network.}
  \label{fig:layout}
\end{center}  \vspace{-8mm}
\end{figure}

In this section, we give the details of the proposed DRL-based algorithm for hybrid beamforming of multi-hop THz communication networks, utilizing the mentioned DDPG algorithm. As seen in Fig. \ref{fig:layout}, the proposed DDPG is composed of the critic neural network
and the actor neural network, which is introduced in detail as follows.

\subsection{Critic and Actor Networks}
As can be seen, both the critic and actor networks  are  comprised of a fully connected DNN, where they share the similar structure, consisted of four layers, i.e., two hidden layers with one input and output layer.  Note that the increase or decrease of the width of the network depends on the actions, but the last output layer is up to the number of users. Based on this, we introduce the batch normalization layer between these two hidden layers with the ReLU activation function. The optimizer used in the critic and actor networks is Adam with learning rate $\mu_c^{(t)}=\lambda_c\mu_c^{(t-1)}$ and $\mu_a^{(t)}=\lambda_a\mu_a^{(t-1)}$, where $\lambda_c$ and $\lambda_a$ represent their the decaying rate.
\subsubsection{Critic Process}
The main objective of critic agent is to evaluate how good a policy is. The input of the critic network are the current environment state and actions generated by the actor network, and outputs the Q-function based on the DDPG. Its learning rate usually is set as smaller to avoid oscillation, but it needs more time to converge. In order to have the negative inputs, the activation function $tanh$  is deployed for the training critic network. To remove the correlation of adjacent states, the input state $\mathbf{S}$ needs the whitening process.

\subsubsection{Actor Process}

The function of the actor network is to learn the current environment based on the proposed DDPG algorithm and outputs the actions to the critic network. Unlike the critic network, the actor needs additional process, i.e., power modular normalization before the output to implementation problems for computing the  $\Delta_{a}q(\mathbf{\theta}_c^{(target)}|s^{(t)},a)$.  The approximate policy gradient might yield the error, and it also could not make sure that we can obtain the optimal solution, but we can minimize the error by using the compatible  features of transition function.

In the actor-critic RL agent, the policy parameters and transition function are updated simultaneously, and $\mathbf{F}$ needs to meet the  power constraint given in (\ref{eq:BD_1}). In order to satisfy this condition, a normalization layer is added at the output of the actor network, in which $Tr \{\mathbf{F}\mathbf{F}^{\mathcal{H}} \} =P_t$.  Noting that, the signal is changed the transmission direction after it goes through the RISs, but its amplitude should be maintained as $\mathbf{\Phi}_i$, $|\mathbf{\Phi}_i(n_i,n_i)|^2=1$ since it does not consume the additional power.

\subsection{Proposed DRL Algorithm }

Before we implement the proposed DRL algorithm, the channel information, $\mathbf{H}_i, i=1,\cdots, I$, $\mathbf{w}_{k}$ and $\mathbf{g}_{k} \forall k$ are collected by the existing methods that are investigated by some previous works \cite{cemaxinying2020,ning2019channel,wei2020ce,wei2020cetcom,xiaojun2020ce}. The channel information and previous actions of $\mathbf{F}^{(t-1)}$ and $\mathbf{\Phi}_i^{(t-1)}, \forall i$ at previous $t-1$ state, the agent obtains the current state $s^{(t)}$ according to the introduction of the section IV.B.1 State. In addition, weight initialization is also a key factor to affect the learning process. The action $\mathbf{F}$ and $\mathbf{\Phi}_i, \forall i$, networks parameters $\mathbf{\theta}_c^{(train)}$ and $\mathbf{\theta}_a^{(train)}$, and replay buffer $\mathcal{M}$ should be initialized before running the algorithm. Furthermore, we also proposed two initialization algorithms, one is based singular value decomposition (SVD), while another one is utilize the max-min SINR method. They are introduced in details as the following IV.C.

The algorithm stops when it converges or reaches the maximum number of iteration steps. The obtained rewards could not be increased with the taking more actions, we think that the output $\mathbf{F}_{opt}, \mathbf{\Phi}_{i,opt}$  are the optimal. Noting that the proposed algorithm might converges the sun-optimal solutions although our objective is to obtain the optimal digital  and analog beamformings. Combing with the previous stated DDPG, the whole proposed DRL-algorithm can be summarized as Algorithm \ref{alg:ALG1} in the following.

\begin{algorithm}[H]
\caption{DRL-based hybrid beamforming design for RIS-based THz Systems}
\label{alg:ALG1}
\textbf{Input:} $\mathbf{w}_{k,\forall k}$, $\mathbf{g}_{k, \forall k},\mathbf{H}_{i,\forall i}$ \\
\textbf{Output:} The optimal $a=\{ \mathbf{F},\mathbf{\Phi}_{i,\forall i} \}$, $Q$ function\\
\textbf{Initialization:} Memory $\mathcal{M}$; parameters $\mathbf{\theta}_c^{(train)}$, $\mathbf{\theta}_a^{(train)}$, $\mathbf{\theta}_c^{(target)}$, $\mathbf{\theta}_a^{(target)}$; \\ beamforming matrices $\mathbf{F}$, $\mathbf{\Phi}_{i,\forall i} $
\begin{algorithmic}[1]
\WHILE{}
\FOR{ \texttt{episode $=0,1,2, \cdots, Z-1$} } 
\STATE Collect and preprocess $\mathbf{w}_{k, \forall k}^{(n)}, \mathbf{g}_{k, \forall k}^{(n)},\mathbf{H}_{i,\forall i}^{(n)}$ for the $n^{th}$ episode to obtain the first state $s^{(0)}$
\FOR{\texttt{t=$0,1,2, \cdots, T-1$}}
\STATE Update action $a^{(t)}=\{\mathbf{F}^{(t)},\mathbf{\Phi}_{i,\forall i}^{(t)} \}=\pi(\mathbf{\theta}_a^{(train)})$ from the actor network \\
\STATE Implement the DDPG Algorithm  \ref{alg:ALGddpg} \\
\STATE Update parameters $\mathbf{\theta}_c^{(train)}$, $\mathbf{\theta}_a^{(train)}$, $\mathbf{\theta}_c^{(target)}$, $\mathbf{\theta}_a^{(target)}$ \\
\STATE Input them to the agent as next state  $s^{(t+1)}$
\ENDFOR
\ENDFOR \\
\textbf{Until:} Convergent or reaches the maximum  iterations.
\ENDWHILE
\end{algorithmic}
\end{algorithm} \vspace{-4mm}

In terms of this proposed algorithm, its state, action, reward and convergence are elaborated in the following.

\subsubsection{State}
The state $s^{(t)}$ is continuous and constructed by the transmit digital beamforming matrix $\mathbf{F}^{(t-1)}$, the analog beamforming matrices $\mathbf{\Phi}_i^{(t-1)}, \forall i$ in the previous $t-1$ time step, and the channel information $\mathbf{H}_i, i=1, \cdots, I$, $\mathbf{w}_k$ and $\mathbf{g}_k, \forall k$. Since the DRL-based  on the TensorFlow platform do not support the complex number inputs, we employ two independent input ports to input the real part and the imaginary part of state $s$ separately. We have the dimension of the state space $D_s=2MK+2\sum_{i=1,\cdots, I}N_i+2MN_1+2\sum_{i=1, \cdots, I-1}N_iN_{i+1}+2KN_I$. We assume that there is no neighboring interference between the different states. To maximize the transmission distance, assume that each state can offer some prior knowledge to DRL agent for selecting the optimal RIS and analog beamforming. The optimal beamforming is related to the channel information and interference to other users. Then, DRL agent can learn the interference patten from the historical date, so that it can infer the future interference at the next time step.

\subsubsection{Action}
Similarly, the action space is also continuous, and  comprised of the digital beamforming matrix $\mathbf{F}$ and analog beamforming matrices  $\mathbf{\Phi}_i, \forall i$. Furthermore, the real and imaginary part of $\mathbf{F}=Re\{\mathbf{F} \}+Im \{ \mathbf{F}\}$ and $\mathbf{\Phi}_i=Re \{\mathbf{\Phi}_i \} +Im \{\mathbf{\Phi}_i \}$ are also separated as two inputs. Its dimension also depends the parameters of communication systems, as $D_a=2MK+2\sum_{i=1,\cdots, I}N_i$.


\subsubsection{Reward}
The instant rewards is affected by two main factors: the contributions to throughput $C(\mathbf{F}^{(t)},\mathbf{\Phi}_i^{(t)}, \mathbf{w}_{k}, \mathbf{g}_{k},\mathbf{H}_{i=1,\cdots, I})$ and the penalty caused by the adjusting the beamforming direction under the prior information, the instantaneous channels $\mathbf{H}_{i=1,\cdots, I}$, $\mathbf{h}_{k}, \forall k$ and the actions  $\mathbf{F}^{(t)}$ and $\mathbf{\Phi}_i^{(t)}$ outputted  from the actor network.

\subsubsection{Convergence}

Furthermore, there are some factors that can affect the convergence. For example, the initialization of action and state parameters plays a key role, which will be introduced in the following. In addition, gradient evolution, learning rate also pose the affect on convergence. The too large or small gradient and learning rate both make a algorithm diverge. We investigate the affect of the learning rate that are shown in simulation section.

\subsection{Action Initialization}
It is well known that, the DRL-based algorithm is not guaranteed to converge to the optimal solution. The convergence speed strongly depends on the initialization of the algorithm. To improve the performance of the proposed algorithm, in this section, we propose two methods to initialize the action $a$.

\subsubsection{Method I}
In this method, the initialized $\mathbf{F}$ and $\mathbf{\Phi}_i, \forall i$ are obtained in two steps.

In this first step, $\mathbf{\Phi}_i, \forall i$ is initialized as any diagonal matrix satisfying $|\mathbf{\Phi}_i(n_i,n_i)|=1$. The initial transmit beamforming matrix is then obtained utilizing zero-forcing (ZF) based algorithm in the second step. The objective of this initialization is to eliminate the co-channel interference between users. Given $\mathbf{\Phi}_i^{(0)}, i=1, \cdots, I$, for the $k^{th}$ user, defining composite channel $\tilde{\mathbf{g}}_k=(\mathbf{g}_k^T\prod_{i=1,\cdots, I}\mathbf{\Phi}_i\mathbf{H}_{i}+\mathbf{w}_k)$ the signal model can be written as
\begin{equation} \label{eq:sys_1aa}
\begin{split}
y_k=\tilde{\mathbf{g}}_{k}^T \mathbf{Fs}+n_k \\
\end{split}
\end{equation}

We construct a matrix as $\tilde{\mathbf{G}}_k=[\tilde{\mathbf{g}}_1, \cdots, \tilde{\mathbf{g}}_{k-1}, \tilde{\mathbf{g}}_{k+1},\cdots,\tilde{\mathbf{g}}_K]^T$. If the $k^{th}$ user transmit signals over the null space of $\tilde{\mathbf{G}}_k$, it will not cause interference to other users. The null space of $\tilde{\mathbf{G}}_k$ can be obtained from singular value decomposition (SVD)
\[\tilde{\mathbf{G}}_k=\tilde{\mathbf{U}}_k\tilde{\mathbf{\Lambda}}_k\tilde{\mathbf{V}}_k^{\mathcal{H}}=\tilde{\mathbf{U}}_k \left [\begin{array}{ccccccc}\tilde{\lambda}_1 &0 &\cdots &0 &0 &\cdots &0 \\
\vdots &\vdots &\cdots &\vdots &\vdots &\cdots &\vdots \\
0 &0 &\cdots &\tilde{\lambda}_{K-1} &0 &\cdots &0 \end{array} \right] \left [ \begin{array}{cccccc}  \tilde{\mathbf{v}}_1 &\cdots &\tilde{\mathbf{v}}_{K-1} &\tilde{\mathbf{v}}_{K} &\cdots &\tilde{\mathbf{v}}_{M} \end{array} \right]^{\mathcal{H}} \]
The null space of $\tilde{\mathbf{G}}_k$ is given by $\tilde{\mathbf{v}}_{K}, \cdots, \tilde{\mathbf{v}}_{M}$. It can be seen that, if the $k^{th}$ user transmits over these null spaces, it will not cause co-channel interference to other users. To guarantee the existence of null spaces, it is required that $M \geq K$. In general, in this method, the transmit beamforming vector for the $k^{th}$ user can be initialized as $\mathbf{f}_k^{(ini)}=\tilde{\mathbf{v}}_i, i=K, \cdots, M$.

\subsubsection{Method II}
In this method, the purpose of the initialization is to maximize the minimum SINR, given the phase shift matrix $\mathbf{\Phi}_i, i=1, \cdots, I$. The optimization problem is formulated as
\begin{equation} \label{eq:BD_33a}
\begin{split}
&\max_{\mathbf{F}} \min_{k=1,\cdots, K} \rho_k(\mathbf{\tilde{\mathbf{g}}}_{k}, \mathbf{F}) \\
& \; \textrm{s.t.} \;\;  tr\{\mathbf{F}\mathbf{F}^{\mathcal{H}} \} \leq P_t \\
& \;\;\;\quad\;\; |\phi_{in_i}|=1\;\forall n_i=1,2,\ldots,N_i,\\
\end{split}
\end{equation}

In \cite{Ref18}, it is shown that, for the $k^{th}$ user, the optimal transmit beamforming vector solving (\ref{eq:BD_33a}) is given by
\begin{equation} \label{eq:BD_33b}
\begin{split}
&\mathbf{F}_k^{(ini)}=\sqrt{p_k^{(ini)}}\frac{\tilde{\mathbf{v}}_k^{(ini)}}{\| \tilde{\mathbf{v}}_k^{(ini)}\|}\\
&\tilde{\mathbf{v}}_k^{(ini)}=\bigg (\sum_{l \neq k}^K q_l^{(ini)}\tilde{\mathbf{g}}_l\tilde{\mathbf{g}}_l^{\mathcal{H}} +\frac{1}{\rho_0}  \bigg )^{-1} \tilde{\mathbf{g}}_k
\end{split}
\end{equation}
where $\rho_0=\frac{P_t}{K\sigma_n^2}$ is the SNR per user. $q_l^{(ini)},l=1, \cdots, K$ is obtained as the unique positive solution to the fixed point equations
\begin{equation} \label{eq:BD_33c}
\begin{split}
q_k^{(ini)}=\frac{\eta^{(ini)}}{\frac{1}{K}\tilde{\mathbf{g}}_k^{\mathcal{H}} \bigg ( \sum_{l \neq k}^K \frac{q_l^{(ini)}}{K}\tilde{\mathbf{g}}_l\tilde{\mathbf{g}}_l^{\mathcal{H}} +\frac{1}{\rho_0}  \bigg )^{-1}\tilde{\mathbf{g}}_k}
\end{split}
\end{equation}
where $\eta^{(ini)}$ is the minimum SINR given by
\begin{equation}\label{eq:BD_34}
\begin{split}
\eta^{(ini)}=\frac{P_t}{\sum_{k=1}^K \bigg ( \tilde{\mathbf{g}}_k^{\mathcal{H}} \bigg ( \sum_{l \neq k}^K q_l^{(ini)}\tilde{\mathbf{g}}_l\tilde{\mathbf{g}}_l^{\mathcal{H}} +\frac{1}{\rho_0}  \bigg )^{-1}\tilde{\mathbf{g}}_k \bigg )^{-1}}
\end{split}
\end{equation}

The power coefficients $p_k^{(ini)}, k=1,\cdots, K$ satisfy the following equations
\begin{equation}\label{eq:BD_35}
\begin{split}
\xi_1^{(ini)}=\xi_2^{(ini)}=,\cdots,=\xi_K=\eta^{(ini)}
\end{split}
\end{equation}
with
\begin{equation}\label{eq:BD_36}
\begin{split}
\xi_k^{(ini)}=\frac{p_k^{(ini)}\frac{|\tilde{\mathbf{g}}_k^{\mathcal{H}}\mathbf{v}_k^{(ini)}|^2}{\| \mathbf{v}_k^{(ini)}\|^2}}{\sum_{l \neq k}^K p_l^{(ini)}\frac{|\tilde{\mathbf{g}}_k^{\mathcal{H}}\mathbf{v}_l^{(ini)}|^2}{\| \mathbf{v}_l^{(ini)}\|^2}+\frac{1}{\rho_0}}
\end{split}
\end{equation}

From the above conditions, we have $\mathbf{p}^{(ini)}=[p_1^{(ini)}, \cdots, p_K^{(ini)}]$
\begin{equation}\label{eq:BD_37}
\begin{split}
\mathbf{p}^{(ini)}=\frac{\eta^{(ini)}}{\rho_0}\big (\mathbf{I}_K-\eta^{(ini)}\mathbf{\Gamma}\mathbf{F}_0  \big)^{-1}\mathbf{\Gamma}\mathbf{1}_K
\end{split}
\end{equation}
where $\mathbf{I}_K$ is an identity, while $\mathbf{1}_K$ is a all one column vector with dimension $K \times K$ and $K \times 1$
\begin{equation}\label{eq:BD_38}
\begin{split}
\mathbf{\Gamma}=diag \{\frac{\| \mathbf{v}_1^{(ini)}\|^2}{|\tilde{\mathbf{g}}_k^{\mathcal{H}}\mathbf{v}_1^{(ini)}|^2}, \cdots,  \frac{\| \mathbf{v}_K^{(ini)}\|^2}{|\tilde{\mathbf{g}}_k^{\mathcal{H}}\mathbf{v}_K^{(ini)}|^2}\}
\end{split}
\end{equation}

The entries of matrix $\mathbf{F}_0, \in \mathbb{C}^{(K \times K)}$ are given by
\[ \mathbf{F}_0(k,l) = \left\{ \begin{array}{ll}
         0 & \mbox{if $k = l$};\\
        \frac{|\tilde{\mathbf{g}}_k^{\mathcal{H}}\mathbf{v}_K^{(ini)}|^2}{\| \mathbf{v}_K^{(ini)}\|^2} &\mbox{if $ k\neq l$}.\end{array} \right. \]

Since the fix point equations (\ref{eq:BD_33c}) cannot be solved directly to obtain the $q_k$, an iterative algorithm that is utilized to obtain the optimal $\mathbf{F}^{(ini)}$ is shown in the following Algorithm \ref{alg:ALG2}.

\begin{algorithm}[H]
\caption{Initialization of Action based on max-min SINR}
\label{alg:ALG2}
\textbf{Initialization:} initialize $\mathbf{p}^{(ini)}, \mathbf{q}^{(ini)}$ such that $\sum p_{l=1}^K=P_t, \sum q_{l=1}^K=P_t$, initialize $\tilde{\mathbf{v}}_l, l=1, \cdots, K$ such that $\| \tilde{\mathbf{v}}_l\|=1$
\begin{algorithmic}[1]
\STATE Compute dual auxiliary variables:\\
\hspace{\algorithmicindent} $\tilde{\mathbf{q}}_l^{(ini)} = \bigg (\tilde{\mathbf{g}}_k^{\mathcal{H}} \bigg ( \sum_{l \neq k}^K q_l^{(ini)}\tilde{\mathbf{g}}_l\tilde{\mathbf{g}}_l^{\mathcal{H}} +\frac{1}{\rho_0}  \bigg )^{-1}\tilde{\mathbf{g}}_k \bigg )^{-1}, l=1,\cdots, K$\\
\STATE Update dual uplink powers: \\
\hspace{\algorithmicindent} $\mathbf{q}^{(ini)}=\frac{P_t}{\sum_{l=1}^K \tilde{q}_l^{(ini)} \tilde{\mathbf{q}}^{(ini)}}$\\
\STATE Compute auxiliary variables: \\
\hspace{\algorithmicindent} $\tilde{p}_l^{(ini)}=\big(\rho_l(\mathbf{F}^{(ini)}) \big)^{-1}p_l^{(ini)}$ \\
\STATE Update downlink powers: \\
\hspace{\algorithmicindent} $\mathbf{p}^{(ini)}=\frac{P_t}{\sum_{l=1}^K \tilde{p}_l^{(ini)} \tilde{\mathbf{q}}^{(ini)}}$\\
\STATE Compute $\tilde{\mathbf{v}}$:\\
\hspace{\algorithmicindent} $\tilde{\mathbf{v}}_k=\bigg (\bigg \| \bigg (\sum_{l \neq k}^K q_l^{(ini)}\tilde{\mathbf{g}}_l\tilde{\mathbf{g}}_l^{\mathcal{H}} +\frac{1}{\rho_0}  \bigg )^{-1} \tilde{\mathbf{g}}_k \bigg \| \bigg )^{-1}\bigg (\sum_{l \neq k}^K q_l^{(ini)}\tilde{\mathbf{g}}_l\tilde{\mathbf{g}}_l^{\mathcal{H}} +\frac{1}{\rho_0}  \bigg )^{-1} \tilde{\mathbf{g}}_k$ \\
\STATE  \textbf{Until} converge, otherwise repeat steps 1-5.
\end{algorithmic}
\end{algorithm} \vspace{-4mm}
where in Step 3, $\rho_l()$ is given by (\ref{eq:sys_3}) while $\mathbf{F}^{(ini)}$ is given by (\ref{eq:BD_33b}).

\section{Numerical Results}

In this section, we numerically evaluate the performance of the proposed DRL-based algorithm for DRL-based hybrid beamforming for multi-hop multiuser RIS-assisted wireless THz communication networks. We firstly introduce the considered settings for the multi-hop THz communication model, and show some results.

\subsection{Simulation Settings}
In the following simulations, we consider a single cell scenario, where there is only one BS, and many RISs that are randomly deployed in a circular region with the diameter as 100 m. The detailed system model and the simulation scenario are given as:
\subsubsection{System Model}
We employ the proposed hybrid beamforming architecture shown in  Fig. \ref{fig:hybrid}. In particular, The BS has $M=8$ antennas with the same number of RF chains, and $K=\{4,32\}$ mobile users equipped the single antenna and RF chain. To reduce the complexity of deployment and learning, we adopt that all $N=64$ RISs have the same number of elements, i.e., $N_i=\{64, 128\}$ for all $i$, and the spacing between
elements equal to $2\lambda$. The small scale fading of channels $\mathbf{w}_{k, \forall k},\mathbf{g}_{k, \forall k},\mathbf{H}_{i-1,\forall i}$ are set as the Rician fading with the factor $K_{H}=K_{g}=K_{w}=1$ except the special declaration in the following. The large scale fading is calculated by the Friis transmission equation.
The transmission frequency is set as 0.12 THz occupied the fixed 12 GHz bandwidth, and the transmitted power of BS is set as $10$ Watt.

\subsubsection{DRL Settings}

We implemented the proposed algorithm with TensorFlow in a general computer, i.e., i7-8700 CPU, Nvidia Geforce GTX 1080Ti. Without special highlight, the parameter settings of the proposed DRL-based beamforming algorithm are concluded in Table \ref{tab:hyperP}.
\begin{table}
\caption{ Parameters for DRL-based beamforming algorithm} \label{tab:hyperP}
\begin{center} \vspace{-4mm}
\begin{tabular}{ | m{4.5em} | m{20em}| m{5em} | }
\hline
Parameters & Description &  Settings \vspace{1mm}\\
\hline
$\beta$ & Discounted rate of the future reward & 0.99  \vspace{1mm}\\
\hline
$\mu_c$ & Learning rate of training critic network update & 0.001  \vspace{1mm} \\
\hline
$\mu_a$ & Learning rate of training actor network update & 0.001 \vspace{1mm} \\
\hline
$\tau_c$ & Learning rate of target critic network update & 0.001 \vspace{1mm} \\
\hline
$\tau_a$ & Learning rate of target actor network update & 0.001 \vspace{1mm} \\
\hline
$\lambda_c$ & Decaying rate of training critic network update & 0.005  \vspace{1mm}\\
\hline
$\lambda_a$ & Decaying rate of training actor network update & 0.005 \vspace{1mm} \\
\hline
$D$ & Buffer size for experience replay& 100000 \vspace{1mm} \\
\hline
$Z$ & Number of episodes & 5000 \vspace{1mm} \\
\hline
$T$ & Number of steps in each episode & 20000 \vspace{1mm} \\
\hline
$W$ & Number of experiences in the mini-batch & 16 \vspace{1mm} \\
\hline
$U$ & Number of steps synchronizing target network with the training network & 1 \vspace{1mm} \\
\hline
\end{tabular} \vspace{-8mm}
\end{center}
\end{table}

\subsubsection{Benchmarks}

To show the effectiveness of our proposed, three significant cases are selected as benchmarks. The first case is an ideal case, where there is no RISs to assist transmit, i.e., $I=0$, and we employ the full digital zero-forcing beamforming. The second typical benchmark was already investigated in some existing works \cite{chongwentwc2019,ning2019beamforming,Nie2020beamforming,Ref10a}, where there is a just single hop between the BS and each user, and  an alternating optimization method is usually proposed to design the beamforming matrices. Furthermore, it is also a special case of our proposed multi-hop scheme, i.e., $I=1$. Unlike the traditional alternating method, we use the proposed DRL-based method to obtain the beamforming matrices that is also seen as the third benchmark.

\subsection{Comparisons with Benchmarks}
\begin{figure}[!t]
\begin{center}
  \includegraphics[width=11cm]{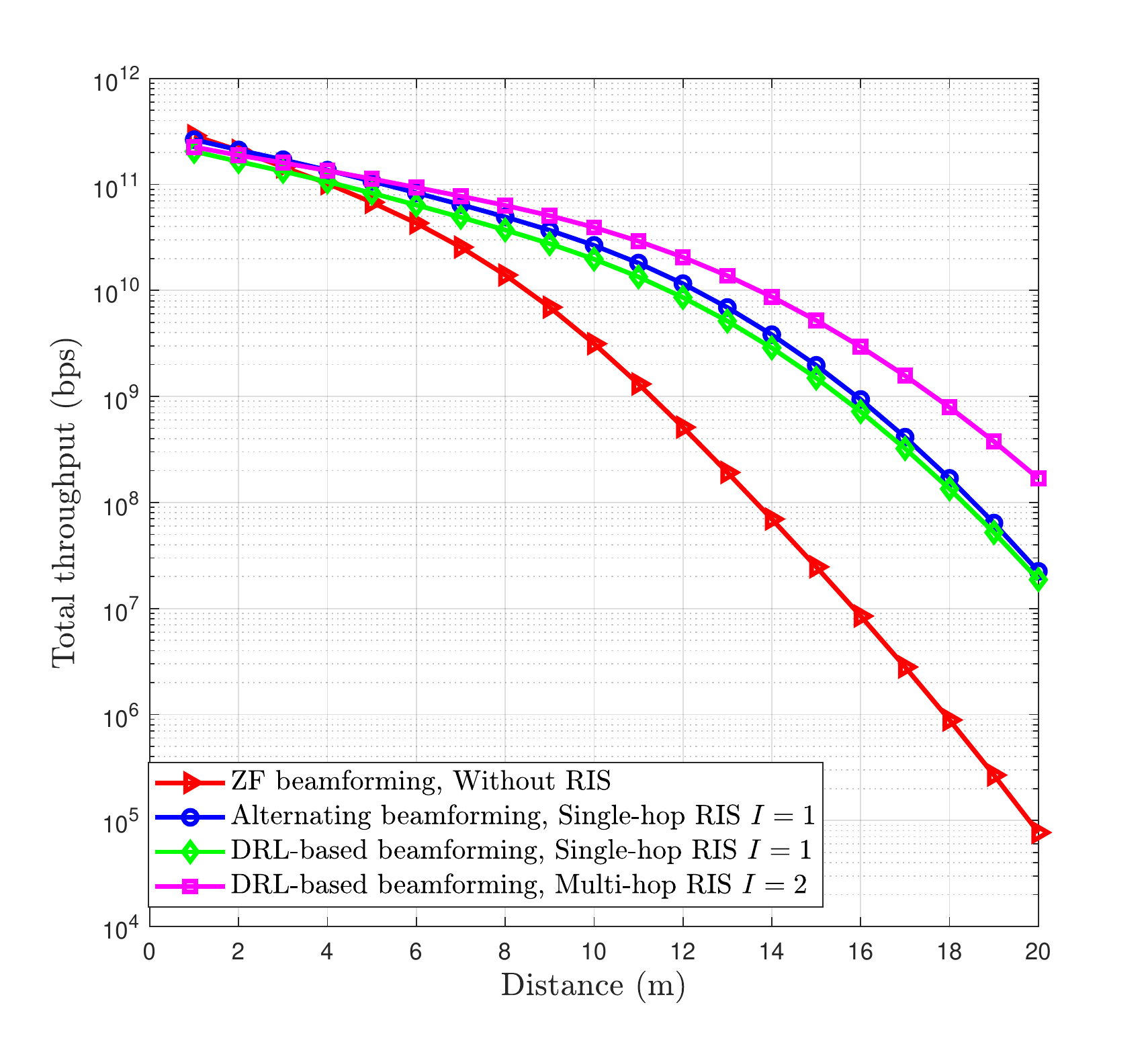} \vspace{-6mm}
  \caption{Total throughput versus transmission distance. We compare  the performance of four schemes, i.e.,  full digital ZF beamforming without RIS, traditional alternating-based beamforming optimization method for the single-hop RIS scenario, the proposed DRL-based beamforming for the single-hop RIS scenario, and the proposed DRL-based beamforming for the multi-hop RIS scenario. }
  \label{fig:comparison}
\end{center} \vspace{-8mm}
\end{figure}

We compare the proposed DRL-based method described in Algorithm 1 for multi-hop RIS-assisted wireless THz communication networks as well as three mentioned benchmarks shown in  Fig. \ref{fig:comparison}. In this simulation, we consider the general system setting of $M=8, N_i=128, K=32$. It shows that the proposed DRL-based multi-hop (i.e., $I=2$) THz communication scheme nearly always obtain the best system throughput compared with the considered three schemes over the whole transmission distance from 1 m to 20 m. In particular, we employ the ideally full digital ZF beamforming for the first benchmark, where does not have the RIS to assist transmission, its throughput drops  fastest with the increase of the transmission distance. For example, under the same throughput 1Gbps, we can see that the proposed DRL-based two-hop scheme obtains around 50\% and 14\% more transmission distances than that of ZF beamforming without RIS and single-hop scheme respectively. What's more, {\color{blue}{this performance gap becomes larger when the transmission distance increases}}. Another interesting point is that the traditional alternating-based method that we adopt is the proposed method in \cite{Ref10a}, as this benchmark can obtain a little better performance than that of the DRL-based beamforming single-hop scheme, but much less than that of the two-hop scheme. These results not only indicate that our proposed scheme can achieve the significant performance gain via multi-hop RISs, but also the DRL-based method can offer the comparable with traditional optimization methods.



\subsection{Impact of System Settings}
\begin{figure}[htbp]
\begin{center}
  \includegraphics[width=11cm]{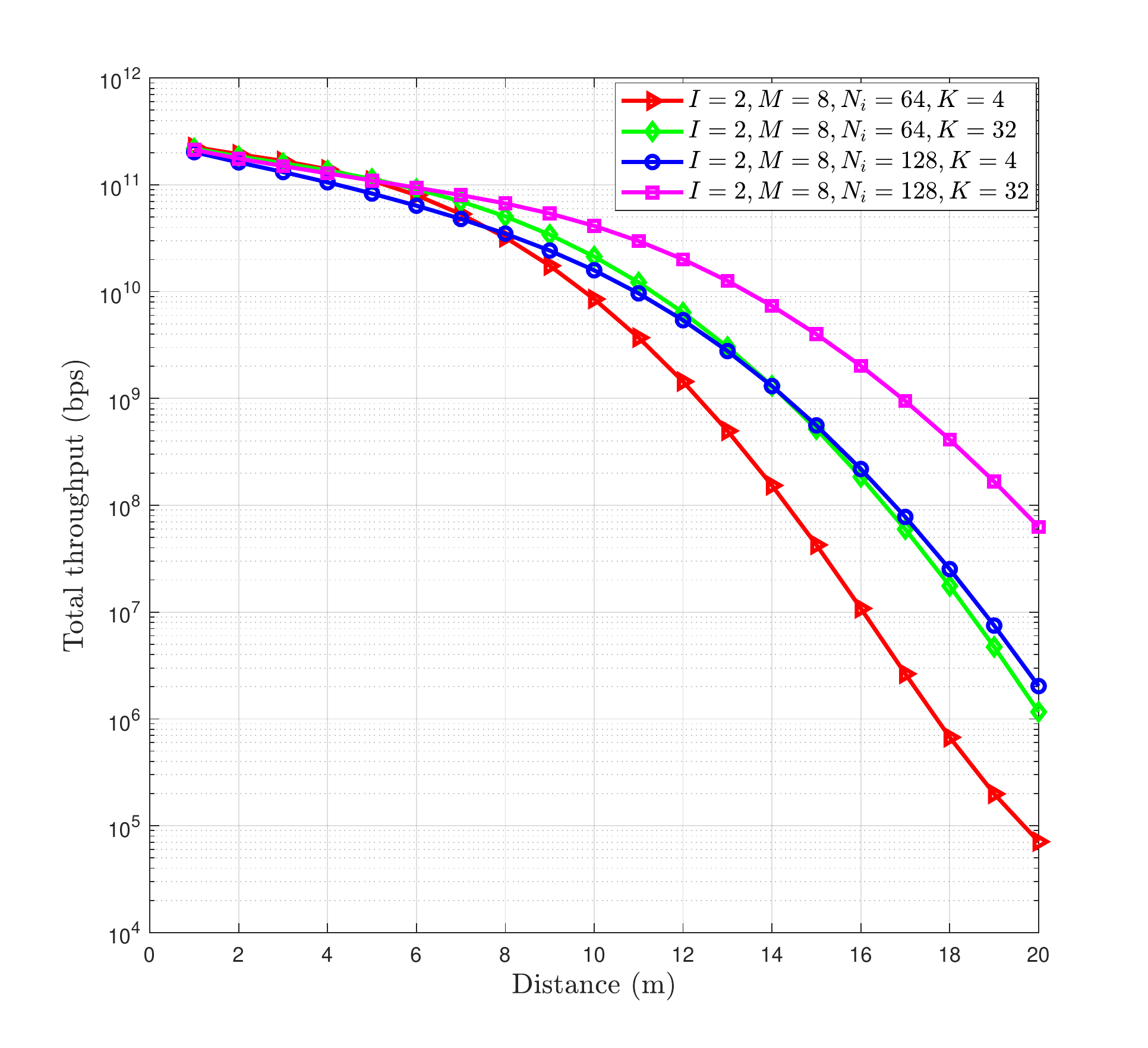} \vspace{-6mm}
  \caption{Total throughput versus transmission distance under different system settings. }
  \label{fig:settings}
\end{center} \vspace{-8mm}
\end{figure}
To further verify the effectiveness of our proposed scheme, we have evaluated its performance under several different system settings, which is shown in Fig. \ref{fig:settings}, where we consider four cases, i.e.,  $N_i=\{64, 128\}$ for all $i$, and $K=\{4, 32\}$ with the common settings of $I=2$ and $M=8$. {\color{blue}{It can be seen that the larger size of RISs and more users result the better total throughput}}. Specifically, when the element number of RISs increases from 64 to 128, the transmission distance improves about 38\%. On the another hand, we increase the number of users, we also obtain the similar results. However, with the increase of transmission distance, the element number of RISs will becomes a key power to combat the path attenuations, e.g., when the transmission distance is larger 15 m.

\begin{figure}[htbp]
\begin{center}
  \includegraphics[width=12cm]{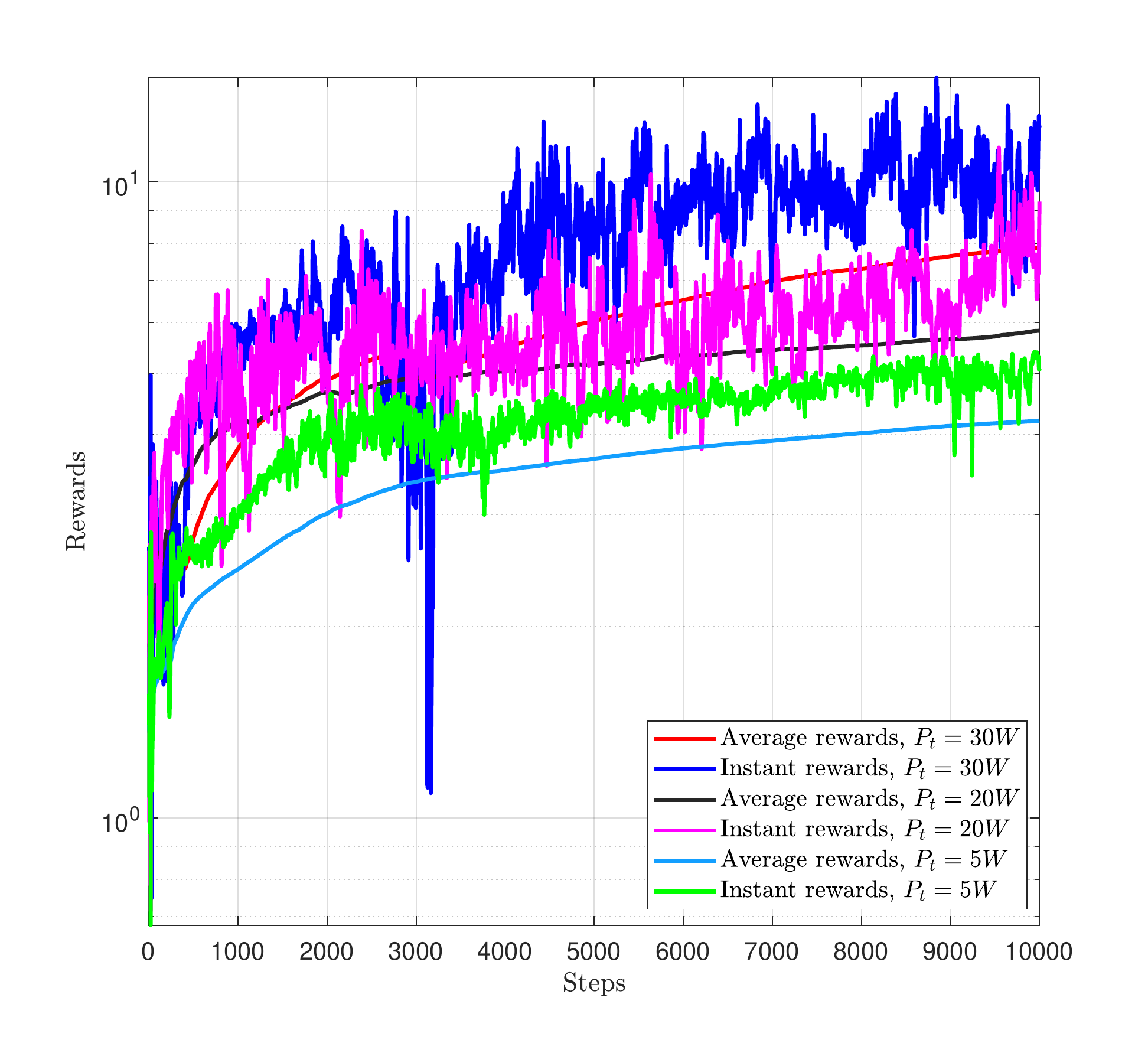} \vspace{-6mm}
  \caption{Rewards versus steps at $P_t=5W$, $P_t=20W$, and $P_t=30W$ respectively. }
  \label{fig:steps5dB} \vspace{-8mm}
\end{center}
\end{figure}

\begin{figure}[htbp]
\begin{center}
  \includegraphics[width=12cm]{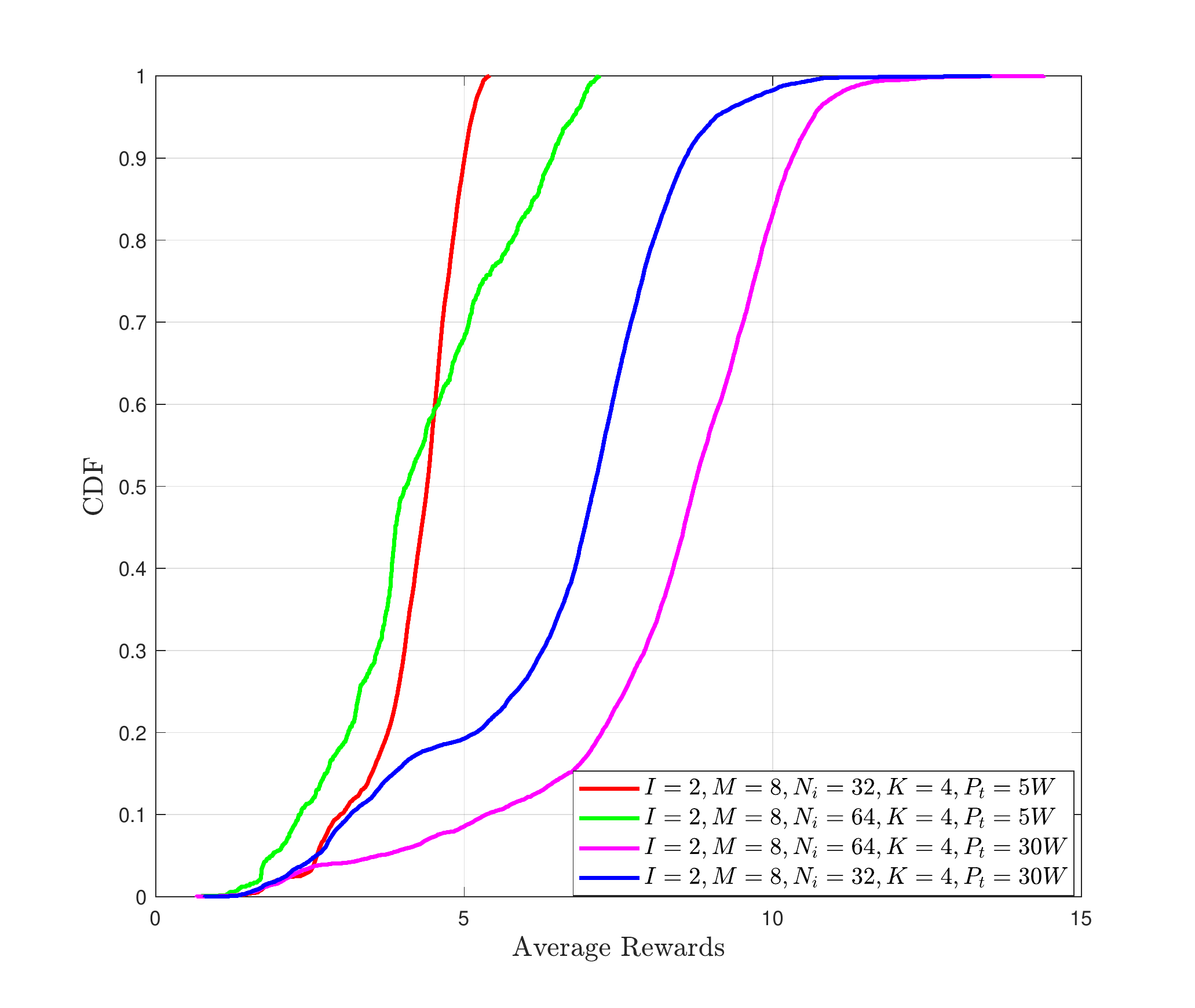} \vspace{-6mm}
  \caption{The cumulative distribution function (CDF) of the average rewards under different system parameters.}
  \label{fig:cdf}
\end{center} \vspace{-8mm}
\end{figure}
The average rewards are calculated by the following equation as
\begin{align} \label{Prob:Stepsize1}
\textrm{average}\_\textrm{reward}(L_i)=\frac{\sum_{l=1}^{L_i} \textrm{reward} (l)}{L_i}, L_i=1,2,...,L,
\end{align}
where $L$ is the maximum steps. The average sum rate as a function of time steps is shown in Fig. \ref{fig:steps5dB} under the setting of $M=8,I=2, N_i=64, K=4$. It can be seen that, the sum rate converges with time step $t$. With the increasing of SNR, instant and average rewards both increase naturally. However, it converges faster at the low transmission power $P_t=5W$ than that of high transmission power $P_t=30W$. This is because the higher transmission power will means the state spaces of instant rewards are larger, which needs to more time to converge the local optimal solution. Based on these results, we also conclude that our proposed DRL-based algorithm can learn from the environment and feed the rewards to the agent to prompt the beamforming matrices $\mathbf{F}$ and $\mathbf{\Phi}_{i,\forall i} $ converging the local optimal.

Furthermore, we also compare the cumulative distribution function (CDF) of the average rewards under different system parameters, i.e., $I=2,M=8$ and $N=\{32,64\},K=\{4,32\},P_t=\{5W,30W\}$, which is shown in Fig. \ref{fig:cdf}. It can be seen that the average rewards will be larger with the increasing of the transmission power and the element number of RISs. In addition, results in this figure also further confirm the part results in Fig.  \ref{fig:settings}, \ref{fig:steps5dB}. Specifically, the transmission power will be a determined factor to improve the average rewards. For the case $P_t=5W$, the element number of RISs has the less improvement compared with the case   $P_t=30W$, even though the average rewards of $N_i=63$ is less than that of $N_i=32$ in some conditions, it is because the lower transmission power plus the multi-user interference will results these week performances of DRL.




\subsection{Impact of the Rician Factor}
\begin{figure}[t]
\begin{center}
  \includegraphics[width=12cm]{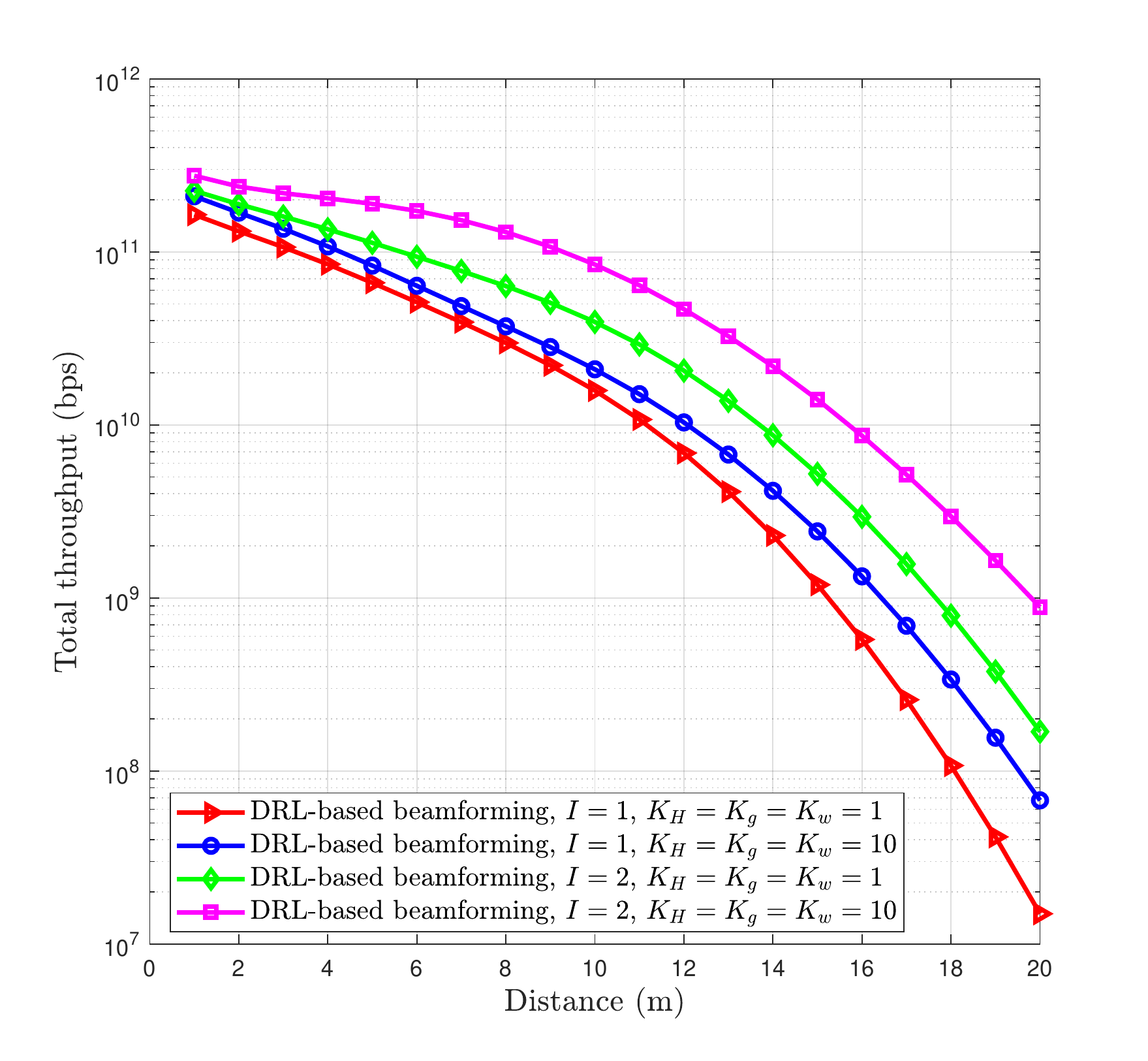} \vspace{-6mm}
  \caption{ Total throughput versus transmission distance. We compare the performance of four cases, i.e.,  $I=1$, $K_{H}=K_{g}=K_{w}=1$,  $I=1$, $K_{H}=K_{g}=K_{w}=10$, $I=2$, $K_{H}=K_{g}=K_{w}=1$  and $I=2$, $K_{H}=K_{g}=K_{w}=10$.} 
  \label{fig:kfactor}
\end{center} \vspace{-8mm}
\end{figure}

In this section, we evaluate the impact of the Rician factor on the total throughput versus transmission distance. We simulate the throughput performance under  four cases, i.e., $I=1$, $K_{H}=K_{g}=K_{w}=1$,  $I=1$, $K_{H}=K_{g}=K_{w}=10$, $I=2$, $K_{H}=K_{g}=K_{w}=1$  and $I=2$, $K_{H}=K_{g}=K_{w}=10$, and the result is shown in Fig. \ref{fig:kfactor}. As shown in the equation (\ref{eq:sys_11}), when the factor $K_{H}=K_{g}=K_{w}=10$, this means that the LoS  component dominates the small scale fading. From this figure, we can see that the fourth case obtain the highest total throughput under the same transmission distance than that of other cases. With the increasing of the Rician factor (i.e., from 1 to 10), the total throughput also increase. This is because the larger Rician factor means that its transmit power is more focused on the LoS path, which reduces the space scatting dissipation, and improves the transmission distance significantly.

\subsection{Impact of the Learning Rate}

\begin{figure}[t]
\begin{center}
  \includegraphics[width=12cm]{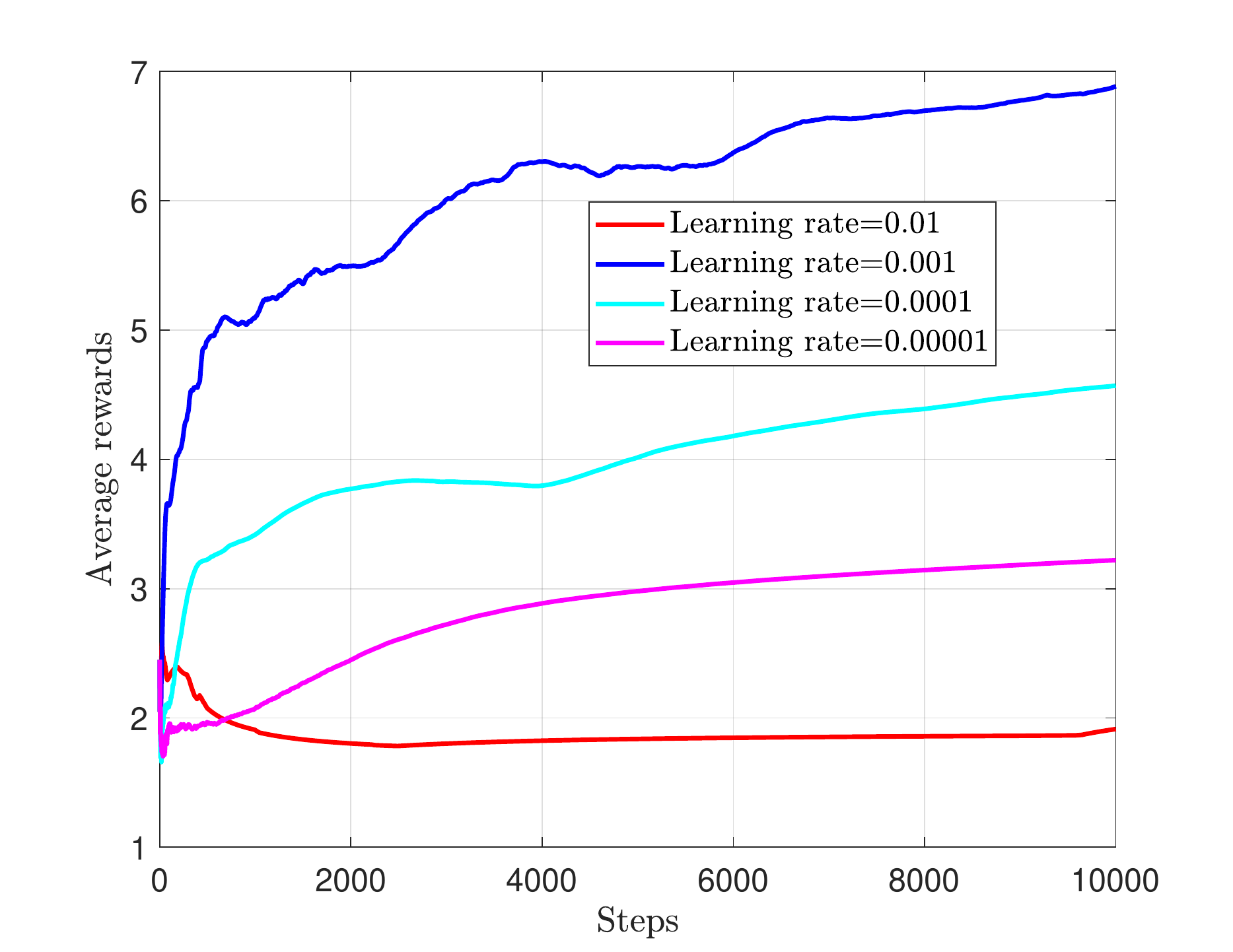} \vspace{-6mm}
  \caption{Average rewards as a function of steps under different learning rates, i.e., \{0.01,0.001,0.0001,0.0001\}.}
  \label{fig:rewardlr}
\end{center} \vspace{-8mm}
\end{figure}


We employ the constant learning rates for our proposed critic-actor networks, and its effects on the convergence and rewards are also investigated as shown in Fig. \ref{fig:rewardlr}, which compares the average rewards under different constant learning rates, i.e., \{0.01,0.001,0.0001,0.0001\}. It should be noted that the learning rate determines the average rewards. Specifically, too large learning (e.g. 0.01) or too small setting (e.g. 0.00001) both results in a lower average rewards, but 0.001 learning rate achieves the best rewards. In contrast, the better rewards performance will increases its convergence time. It is because the better rewards means the larger state space that might need more time to converge the optimal solution.


Finally, we can draw the conclusion that DRL is the complex learning processing, where its performance can be affected by a variety of factors, especially when the environment is fast changing. These factors not only  include the hyper-parameters, learning rate, decaying rate, minibatch size, etc, but also the initialization and system settings. It also should be noted that appropriate settings and tuning the parameters of DRL will be beneficial to improve the performance and reduce the convergence time.

\section{Conclusions}
In this paper, a practical hybrid beamforming architecture for multi-hop multiuser RIS-empowered wireless THz communication networks was proposed, which can effectively combat the severe propagation attenuations and improve the coverage range. Based on this proposed scheme,  a non-convex joint design problem of the digital beamforming and analog beamforming matrices was formulated.  To tackle this NP-hard problem,  a practical DRL-empowered algorithm was proposed, which is a very
early attempt to address this hybrid design problem. To improve the convergence of the proposed DRL algorithm, two methods are presented to initialize the digital beamforming and analog beamforming matrices. Moreover, the proposed DRL-based algorithm has a very standard formulation and low complexity in implementation, without knowledge of explicit model of wireless channels and specific mathematical formulations. It is therefore very easy to be scaled to accommodate various system settings. Simulation results show that our proposed scheme is able to improve 50\% more coverage range of THz communications compared with the considered benchmarks. Furthermore, it is also shown that our proposed DRL-based method is a state-of-the-art method to solve the NP-hard beamforming problem, especially when the signals at RIS-assisted THz communication networks experience multi hops.

\bibliography{multihop_ris}
\bibliographystyle{IEEEtran}

\end{document}